\documentclass[12pt]{article}
\usepackage{graphicx}
\usepackage{url} 
\newcommand{\blind}{0}

\usepackage{amsmath, amsthm, amssymb,bm}  
\usepackage{graphicx, subcaption, caption, float, arydshln}  
\usepackage{algorithm}
\usepackage{algorithmicx}
\usepackage{algpseudocode}

\usepackage{natbib}

\usepackage{xcolor}
\newcommand{\Expect}{\mathbb E}

\newcommand{\Var}{\mathbb V}
\newcommand{\X}{\mathcal X}

\newcommand{\Real}{\mathbb R}
\newcommand{\x}{\bm{x}}
\newcommand{\y}{\bm{y}}
\newcommand{\f}{\bm{f}}
\DeclareMathOperator*{\argmax}{arg\,max}
\DeclareMathOperator*{\argmin}{arg\,min}

\usepackage{xr}
\makeatletter
\newcommand*{\addFileDependency}[1]{
\typeout{(#1)}
\@addtofilelist{#1}
\IfFileExists{#1}{}{\typeout{No file #1.}}
}\makeatother
\newcommand*{\myexternaldocument}[1]{%
\externaldocument{#1}%
\addFileDependency{#1.tex}%
\addFileDependency{#1.aux}%
}
\myexternaldocument{supp}

\begin{document}

\def\spacingset#1{\renewcommand{\baselinestretch}%
{#1}\small\normalsize} \spacingset{1}


\if0\blind
{
  \title{\bf Sequential Designs for Filling Output Spaces}
  \author{Shangkun Wang\hspace{.2cm}\\
    H. Milton Stewart School of Industrial and Systems Engineering,\\ Georgia Institute of Technology\\
    and \\
    Adam P. Generale, Surya R. Kalidindi\\
    George W. Woodruff School of Mechanical Engineering,\\Georgia Institute of Technology\\ and \\
    V. Roshan Joseph\\
    H. Milton Stewart School of Industrial and Systems Engineering,\\ Georgia Institute of Technology
}
  \maketitle
} \fi

\if1\blind
{
  \bigskip
  \bigskip
  \bigskip
  \begin{center}
    {\LARGE\bf Sequential Designs for Filling Output Spaces}
\end{center}
  \medskip
} \fi

\bigskip
\begin{abstract}

Space-filling designs are commonly used in computer experiments to fill the space of inputs so that the input-output relationship can be accurately estimated. However, in certain applications such as inverse design or feature-based modeling, the aim is to fill the response or feature space. In this article, we propose a new experimental design framework that aims to fill the space of the outputs (responses or features). The design is adaptive and model-free, and therefore is expected to be robust to different kinds of modeling choices and input-output relationships. Several examples are given to show the advantages of the proposed method over the traditional input space-filling designs.
\end{abstract}

\noindent%
\vfill

\newpage
\spacingset{1.8} 
\section{Introduction}
\label{sec:intro}

Computer experiments have become an indispensable tool in science and engineering \citep{santner2019design}, such as in rocket engine design \citep{mak2018efficient}, biomedical engineering \citep{striegel2022multifidelity},  and materials design \citep{Iyer2022}. However, with the increasing complexity and resolution of the simulations, the running time of computer experiments is still far from negligible even with the current computational power. Thus, experimental designs that help to gather maximum information with minimum computational budget play a crucial role in computer experiments.

Space-filling designs are widely used as experimental designs for computer experiments \citep{joseph2016space}. Intuitively speaking, a space-filling design tries to place the design points to ``fill" the \textit{input} space  well in the hope that the estimation and prediction of the statistical model based on the experimental data would be accurate. An attractive feature of space-filling designs is that they are robust to modeling assumptions. Thus, they can be efficiently used for fitting a wide variety of models.




In this article, we are interested in experimental designs that produce response values to fill the output spaces. We refer to them as output space-filling designs (OSFD).  Unlike traditional space-filling design (hereafter referred to as input space-filling design (ISFD)), OSFD aims to cover the \textit{output} space well. Generating OSFD may seem like an unusual objective because filling the output space does not guarantee a precise estimation of the input-output relationship. On the other hand, the benefits of filling the input spaces are well known. \cite{johnson1990minimax} have shown that a maximin distance design in the input space would be asymptotically D-optimal for fitting a Gaussian process model and a minimax design would be asymptotically G-optimal (as the correlations tend to zero). However, as noted by \cite{lu2021input}, there are several applications in which filling the output space would be beneficial.


In inverse design problems, the aim is to find  input configurations that will achieve  a specified set of outputs. Consider, for the example, the design of acoustic metasurfaces to achieve given acoustic properties such as the amplitude and phase of transmitted and reflected waves \citep{krishna2022inverse}. The aim is to create several acoustic metasurfaces offline using 3D printing, store them, and pick the best one for a given set of acoustic properties. In other words, the aim is to create a ``lookup-table'' of acoustic metasurface geometries and acoustic properties, where the investigator can quickly identify the geometry based on the set of acoustic properties. This approach will be successful if the acoustic properties in the lookup table is dense, that is, the space of acoustic properties should not have large gaps. Thus, the aim here is to identify the set of geometries that will fill the output space of acoustic properties.

As a second application, consider statistical and machine learning problems involving feature-based modeling. The first step in such problems is to extract ``features'' from the input space. The modeling is then done between the response and features. In this scenario, the feature-output relationship can be accurately estimated if the points in the feature space is space-filling. Thus, the aim is to identify a set of points in the input space so that the points in the feature space are space-filling.  As a real example, consider the crystal structure prediction problem described in \cite{krishna2022crty}. The input space is the Cartesian coordinates of the atomic configurations of a single crystal structure, and the output is the potential energy computed using Density Functional Theory (DFT). However, since the potential energy is invariant to translational, rotational, and permutational operations of the atoms, the Cartesian cordinate system is not suitable for model building. Therefore, the Cartesian coordinates are converted to a set of features using AGNI (Adaptive, Generalizable and Neighborhood Informed) fingerprinting \citep{batra2019general}. AGNI fingerprinting is fast compared to DFT computations, but has non-negligible cost which makes developing space-filling points in the feature space a difficult task.



It is much more challenging to generate space-filling points in the output (response, feature, etc.) space as compared to the input space. ISFD is generated in a known experimental region, usually a hypercube, whereas OSFD aims to fill an unknown region with unknown boundaries. A naive approach of creating a large set of candidate points in the output space to choose a set of space-filling points would not be feasible in either of the two scenarios: (1) high cost of evaluating the input-output function and (2) the region in the input space to cover the output space is small relative to the whole input space. This suggests that a sequential design is the only viable option, as we can learn the ``active regions'' in the input space  gradually, and fill-in the output space with as few function evaluations as possible. Developing such a sequential design is the main aim of this article. 


Sequential designs, also known as active learning, is widely used in statistics and machine learning for dealing with expensive black-box functions. Most of these are ``model-based'' designs and use Gaussian process modeling extensively \citep[Ch.6]{gramacy2020surrogates}. However, Gaussian process modeling has a high cost for training, which can be appreciable in several applications. Therefore, we need new sequential design methods that is fast and efficient to fill-in the output spaces. 



The article is outlined as follows. Section \ref{background} begins by introducing some notations used in this article and then reviews the traditional ISFDs and related works on OSFDs. Section \ref{method} presents the definition of minimax output space-filling design and proposes efficient algorithms to generate such designs. Section \ref{results} demonstrates the performance of the proposed algorithms using three simulation studies. Section \ref{application} illustrates the application of OSFD on inverse design and feature-based modeling. Section \ref{conclusion} concludes the article with some final remarks.

\section{Background}\label{background}
Denote the input space by $\mathcal{X} \subseteq {\Real} ^p$, output space by $\mathcal{Y}\subseteq {\Real} ^q$, and let the mapping from input space to output space be $\f:\mathcal{X}\rightarrow \mathcal{Y}$. Typically $\f$ is a black-box computer code that is expensive to evaluate. Denote a design of size $n\in \mathbb{N}$ by $\mathcal D_n=\{\x_i\in\mathcal{X},i=1,\dots,n\}$. The corresponding points in the output space is denoted by $\mathcal{M}_n=\f(\mathcal{D}_n)=\{\y_i: \y_i=\f(\x_i),\x_i\in\mathcal{D}_n,i=1,\dots,n\}$. Our goal is to find a design $\mathcal{D}_n$ such that $\mathcal{M}_n$ is space-filling in $\mathcal Y$. Before we formally define what is ``space-filling" in $\mathcal Y$, we first review the traditional input space-filling design (ISFD).

\subsection {Space-filling Design}
From a geometric point of view, there are two commonly used space-filling design schemes: maximin distance design and minimax distance design \citep{johnson1990minimax}. Let $d_x$ be a metric on $\Real^p$. Then the maximin distance design maximizes the following criterion:
\begin{equation}
    \phi_{\text{Mm}}(\mathcal{D}_n)=\min_{\x_i,\x_j\in\mathcal{D}_n;i\neq j} d_x(\x_i,\x_j).
\end{equation}
That is, it places design points such that the minimum distance between any two points is as large as possible. Minimax distance design, on the other hand, tries to minimize the maximum distance from all the points $x\in \mathcal{X}$ to their closest neighbor in $\mathcal{D}_n$, which is obtained by minimizing the following criterion:
\begin{equation}\label{eq:minimax_x}
    \phi_{\text{mM}}(\mathcal{D}_n)=\max_{x\in\mathcal{X}}\min_i d_x(\x,\x_i).
\end{equation}
This criterion is also known as the fill distance \citep{fasshauer2007meshfree}, which will be used throughout this article. Maximin and minimax distance designs may not have good projection properties and therefore, they are  combined with Latin hypercube designs (LHD) to improve their one-dimensional projections \citep{morris1995exploratory}. We refer the readers to \cite{joseph2016space} for a detailed review of the vast literature on ISFD. These designs  allow for a careful exploration of the experimental region by making sure that no part of the input space is left out. This property makes these designs model-free and therefore, they enable the experimenter to fit a wide variety of statistical and machine learning models to the data and make predictions. 



\subsection {Related works}

The literature on output space-filling design (OSFD) is scarce. \cite{rhee2017space} seems to be the first work that discussed about space-filling designs for output spaces. However, their goal is more closely aligned with uniform sampling on a manifold rather than generating an experimental design. To generate $n$ uniform points in the output space, they start with $n$ random samples in the input space and then improve them through weighting and resampling. However, in the context of design of experiments, augmentation of the design points makes more sense than resampling. In this article, we will develop a sequential design strategy that adds points one-at-a-time to the existing set of points, thereby obtaining a space-filling design with minimum number of function evaluations. Non-uniformity of the points in the output space is not a concern at all for us as long as the points can fill-in the output space.

\cite{lu2021input} recently proposed a design strategy that simultaneously achieve space-fillingness in both input and output spaces using Pareto front optimization. They assume that the input-output relationship is known and cheap to evaluate, which is quite different from the problem we tackle in this article. \cite{lu2021input} also proposed a two-stage approach to deal with the unknown input-output relationship by first using an ISFD to estimate the relationship and then using the estimated model to perform the Pareto front optimization. Their second stage design can suffer if the estimated model is wrong from the first stage. In contrast, we develop a fully sequential model-robust design strategy to construct the OSFD.

Our work is motivated by the two applications briefly discussed in Section 1: inverse design of acoustic metasurfaces \citep{krishna2022inverse} and crystal structure prediction \citep{krishna2022crty}. The authors develop design strategies specific to those two applications. In contrast, the design strategy developed here is more general, efficient, and broadly applicable.

\section{Output Space-Filling Design}\label{method}

\subsection{Mathematical Formulation}
Similar to the traditional minimax distance design, here we quantify the space-fillingness of the design output using the minimax distance. Therefore, our objective is to minimize
\begin{equation}\label{eq:minimax_y}
    \phi_{\text{mM}}(\mathcal{M}_n)=\max_{\y\in\mathcal{Y}}\inf_{\y_i\in \f(\mathcal{D}_n)} d_y(\y,\y_i),
\end{equation}
with respect to $\mathcal{D}_n$, where $d_y$ is a metric defined on $\Real^q$. We use Euclidean distance as the metric unless otherwise mentioned. We call the minimizer of the foregoing objective function a minimax output space-filling design. The only difference between  \eqref{eq:minimax_x} and  \eqref{eq:minimax_y} is the domain to which we apply the minimax distance criterion. However, due to the unknown mapping $\f$ and unknown  output space $\mathcal{Y}$ in  \eqref{eq:minimax_y}, obtaining a high-quality space-filling design in the output space is almost impossible and the only hope is to develop numerical algorithms that can at least approximate this idealized aim as close as possible. 

\begin{figure}[ht]
    \centering
    \includegraphics[width=1\textwidth]{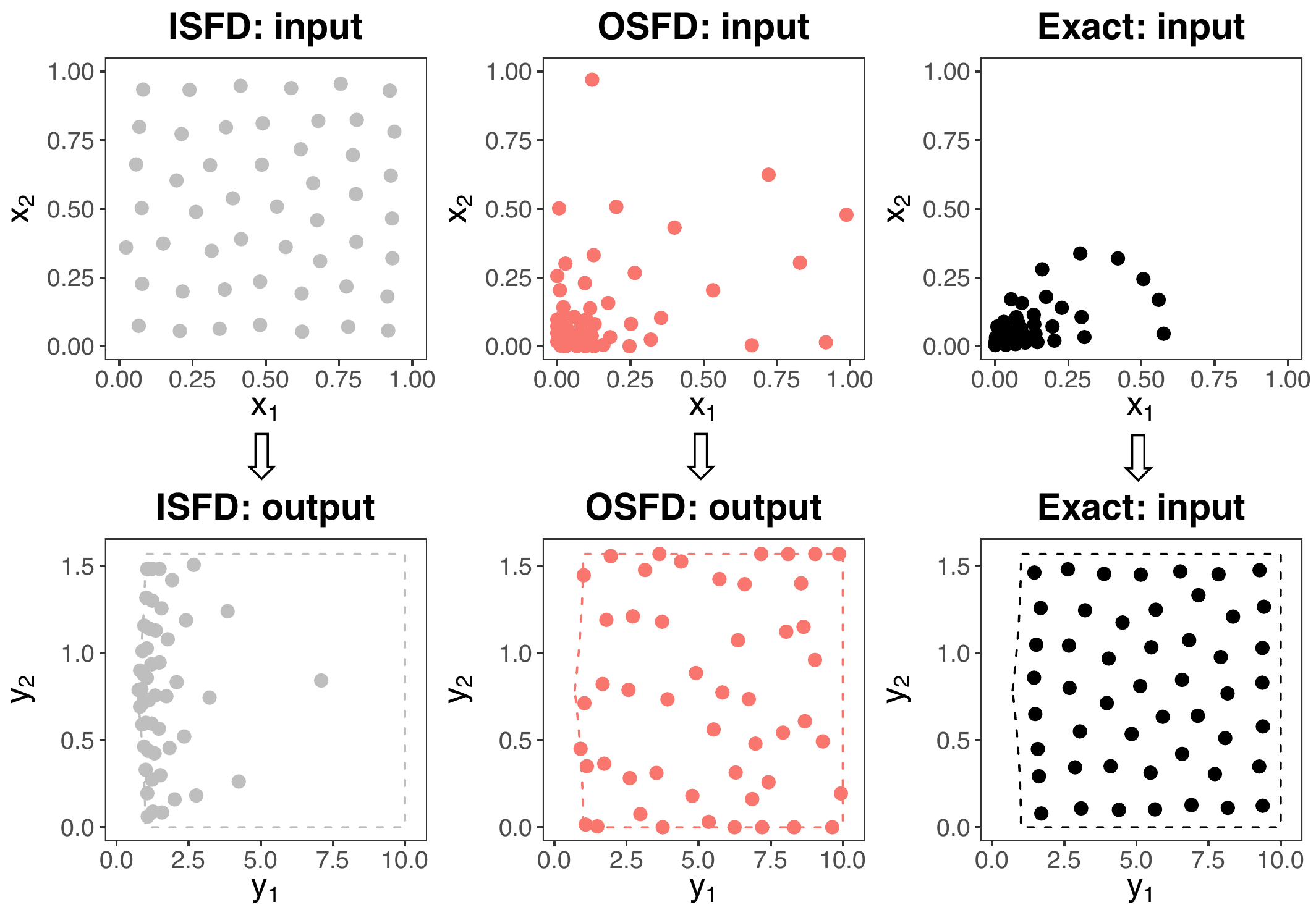}
    \caption{Design points and outputs for the inverse radius function in \eqref{eq:ir} with $\epsilon=0.1$. Input space-filling design (left); output space-filling design by OSFD-greedy algorithm (middle); and the exact solution of the minimax OSFD (right). Design size is 50. The output space is enclosed by the dashed line. The initial design for the OSFD-greedy algorithm is a random LHD of size 5.}
    \label{fig:intro}
\end{figure}

For illustration, consider a simple `inverse-radius' function $\f_{\text{ir}}$ that maps $[0,1]^2$ to a subset in $\Real^2$:
\begin{equation}\label{eq:ir}
    \f_{\text{ir}}(x_1,x_2)=\left(\frac{1}{\sqrt{x_1^2+x_2^2+\epsilon^2}},\arctan{\frac{x_2}{x_1}}\right).
\end{equation}
This function has large gradient near the origin and is flat elsewhere, with $\epsilon$ controlling the variation. As shown in the left panels in Figure \ref{fig:intro}, if we use the traditional minimax design, the output points will congregate at the left side of the output space and leave most of $\mathcal{Y}$ unexplored. Since $\f_{\text{ir}}$ is a bijection, to get the theoretical minimax output space-filling design, we can first construct a minimax design in output space, which we denote as $\mathcal{M}_n^*=\arg\min_{\mathcal{M}_n}\phi_\text{mM}(\mathcal{M}_n)$, and then map it back to the input space to obtain $\mathcal{D}_n^*=\f_{\text{ir}}^{-1}(\mathcal{M}_n^*)$. Because the output space is irregularly shaped, here we use the minimax clustering with particle swarm optimization (mM-PSO) algorithm of \cite{mak2018minimax} to generate the minimax points in the region enclosed by the dashed line on the right panel of Figure \ref{fig:intro}. In this case, the output space is covered uniformly and the design points in the input space exploits the ``interesting" region where the function $\f_{\text{ir}}$ has large variation. In practice, though, we have no knowledge of the output space and $\f_{\text{ir}}$ is a black-box function that can only be evaluated in the forward direction. Therefore, this exact optimal design is impossible to attain in practice.  
Interestingly, our sequential output space-filling design algorithm discussed in the next subsection can find a compromise between the two aforementioned cases, recovering most part of the output space while exploring the input space well (see the middle panels of Figure \ref{fig:intro}).


\subsection{A Sequential Design Algorithm}


Our algorithm consists of two steps: (i) find the largest gap in the output space and (ii) perturb the corresponding input point to generate a new design point in the input space. These two steps will be continued until the largest gap in the output space is below a specified threshold or when the budget is run out. We will now describe the two steps in detail.

\subsubsection{Gap identification}
Suppose we already have $m$ points in the input and output spaces: $(\mathcal{D}_m,\mathcal{M}_m)$. The first step is to identify the largest gap in the output space. For this purpose, we define the local fill distance around each point $\y_i$ as
\begin{equation}\label{eq:loc_fill}
    h_i = \max_{\y\in V_i^{out}} d_y(\y,\y_i),
\end{equation}
where $V_i^{out}$ is the Voronoi region around $\y_i$ given by
\begin{equation}\label{voro}
    V_i^{out} =\{\y\in \mathcal{Y}, d_y(\y,\y_i)\leq d_y(\y,\y_j),\;  \forall \y_j\in\mathcal{M}_m\neq \y_i\},
\end{equation}
for $i=1,\ldots,m$. Note that $h_i$ and $V^{out}_i$ depend on the current outputs $\mathcal{M}_m$. For notational compactness, we have dropped their dependence on $\mathcal{M}_m$ as long as it is clear that these quantities will change as more points are added to the design. 
It is easy to see that the fill distance of $\mathcal{M}_m$ is
\[\phi_{\text{mM}}(\mathcal{M}_m)=\max_{i=1:m} h_i.\]
The index of the point corresponding to the largest gap in the output space is given by
\begin{equation}\label{eq:istar}
    i^*=\argmax_{i=1:m} h_i.
\end{equation}
With the point of largest local fill distance identified, we can perturb $\x_{i^*}$ in the input space. However, the evaluation of local fill distance requires the knowledge of the true $\mathcal{Y}$, which is actually unknown beforehand. This renders a direct segmentation of the output space into Voronoi regions (Eq. \ref{voro}) not feasible. Therefore, based on $\mathcal{M}_m$, we first generate a set  of points $\mathcal{A}$ to approximate $\mathcal{Y}$ as follows. 

\begin{figure*}
        \centering
        \begin{subfigure}[b]{0.4\textwidth}
            \centering
            \includegraphics[width=\textwidth]{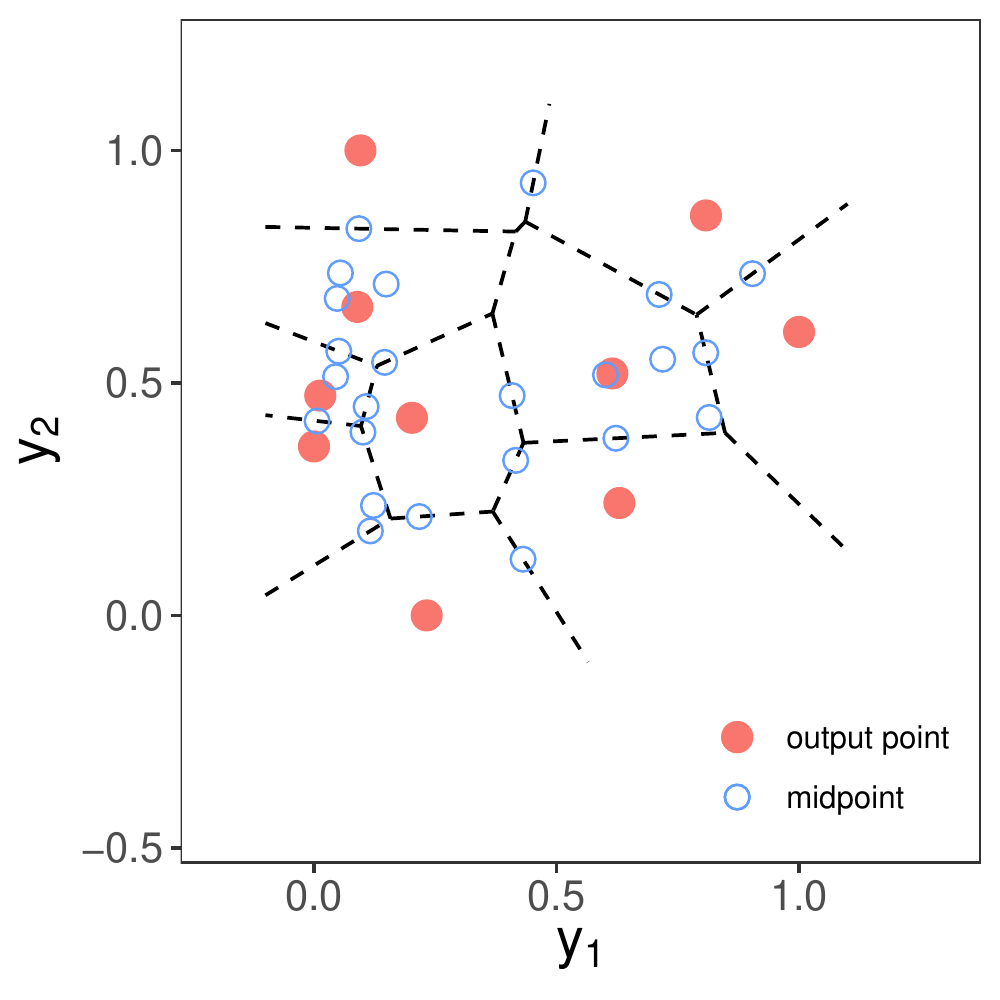}
            \caption
            {{\small $\mathcal{A}_1$}}    
            \label{fig:A1}
        \end{subfigure}
        \begin{subfigure}[b]{0.4\textwidth}  
            \centering 
            \includegraphics[width=\textwidth]{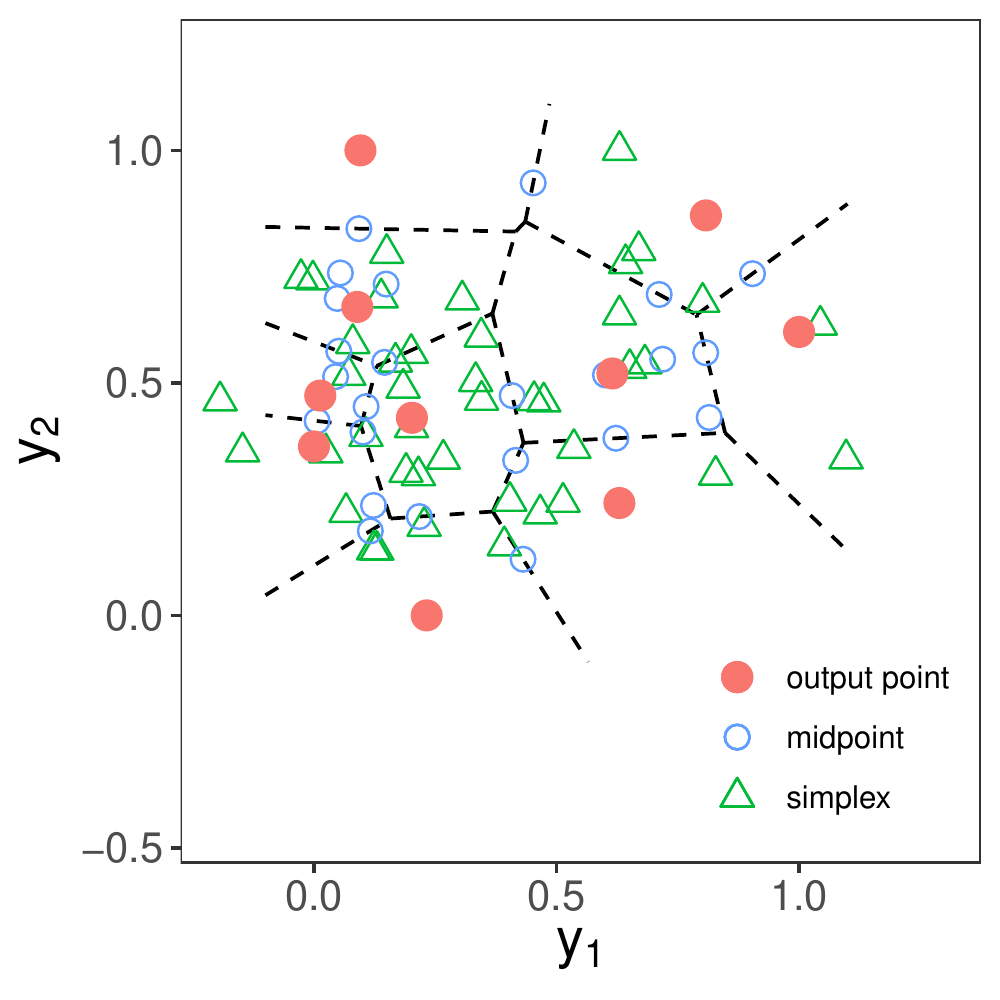}
            \caption
            {{\small $\mathcal{A}_1\cup\mathcal{A}_2$}}    
            \label{fig:A2}
        \end{subfigure}
        \begin{subfigure}[b]{0.4\textwidth}   
            \centering 
            \includegraphics[width=\textwidth]{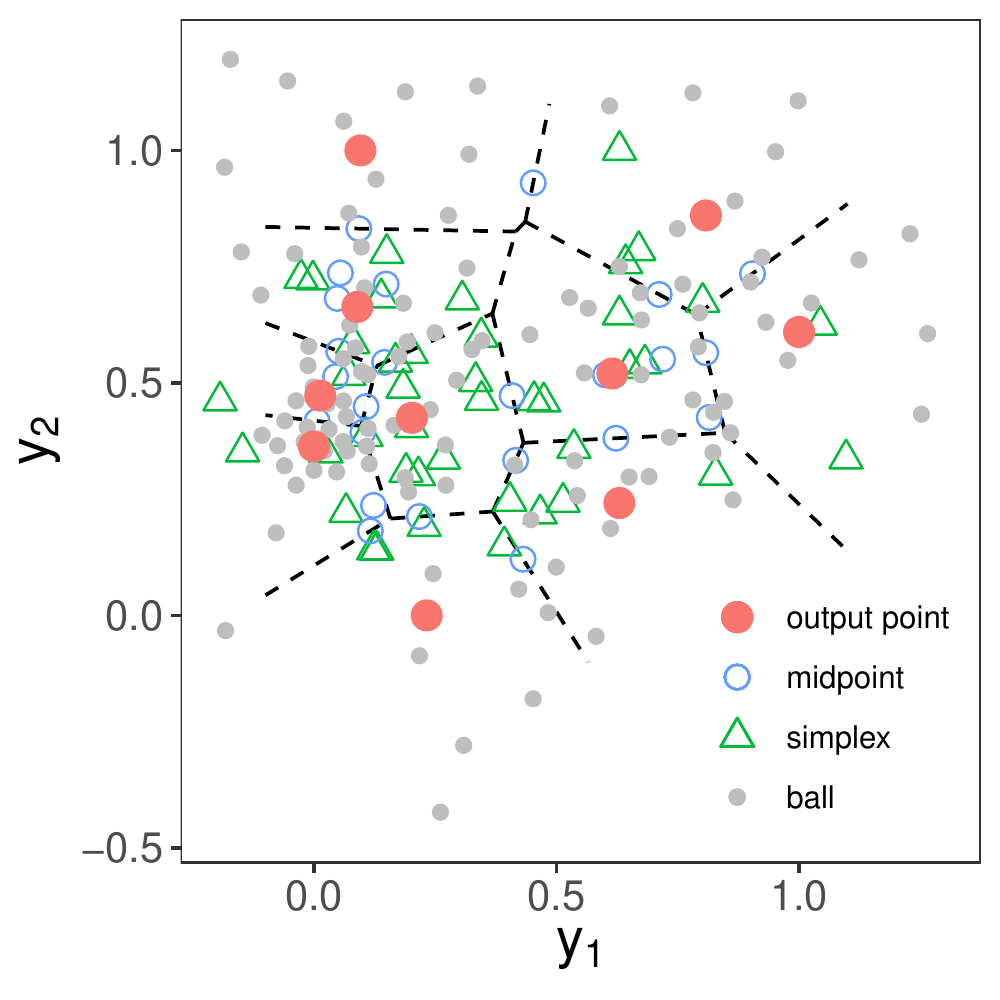}
            \caption
            {{\small$\mathcal{A}_1\cup\mathcal{A}_2\cup\mathcal{A}_3$}}
            \label{fig:A3}
        \end{subfigure}
        \begin{subfigure}[b]{0.4\textwidth}   
            \centering 
            \includegraphics[width=\textwidth]{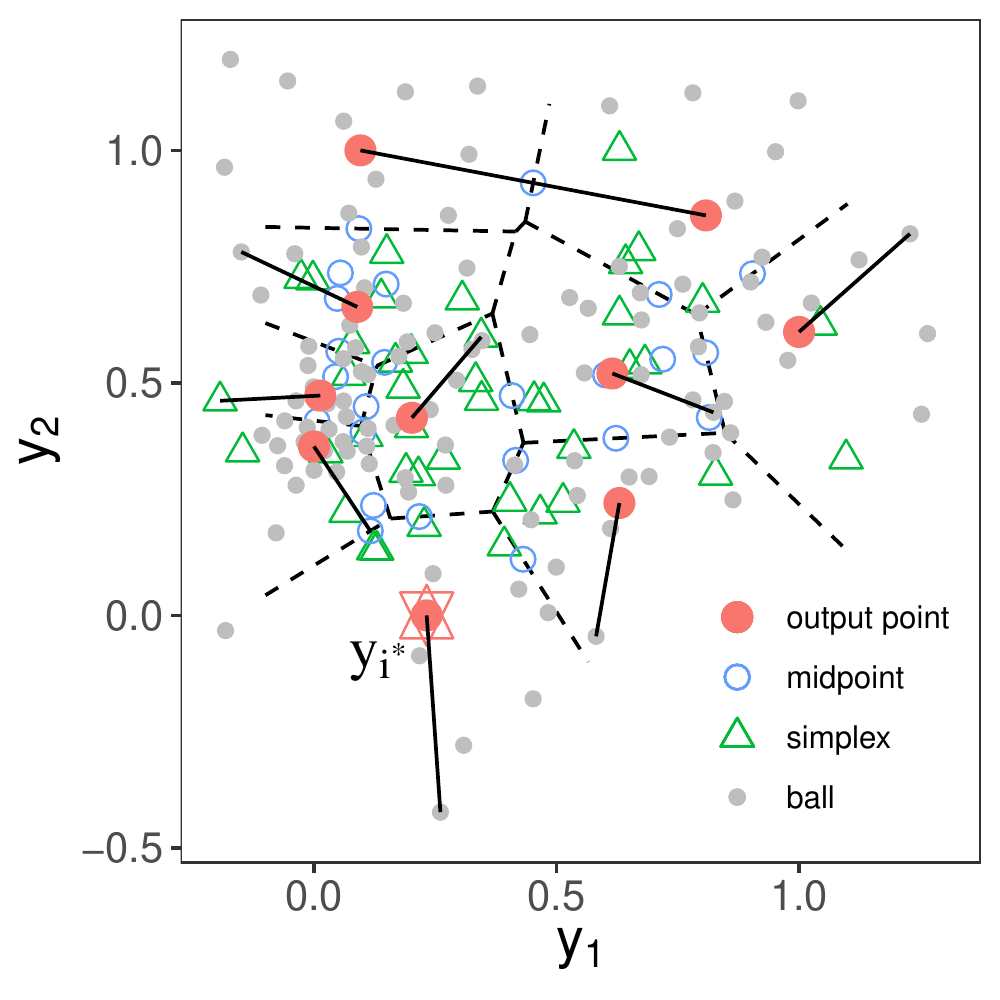}
            \caption[local fill-distance]%
            {{\small Local fill distance}}   
            \label{fig:A4}
        \end{subfigure}
        \caption{Construction of approximating points $\mathcal{A}$: Red points are the scaled output points $\mathcal{M}_{10}$. Dashed lines divide the output region into 10 Voronoi cells, within which each output point is connected to its furthest approximating point. The star in the last figure is the design output $\y_{i^*}$with largest local fill distance.}
        \label{fig:filldist}
    \end{figure*}

The approximating point set is comprised of three parts. The first part $\mathcal{A}_1$ is generated by constructing a $(p\wedge q+1)$-dimensional simplex by connecting each $\y_i \in \mathcal{M}_m$ and its $p\wedge q$ nearest neighbors $N^{p\wedge q}(\y_i)=\{\y_i^{(l)}\in\mathcal{M}_m:\y_i^{(l)}\;\text{is the $l$th nearest neighbor of $\y_i$},l=1,\dots,p\wedge q\}$ and then finding the centroid:
\begin{equation}\label{eq:centroid}
                \bm c = \frac{1}{p\wedge q+1}\sum_{l=0}^{p\wedge q} \y_i^{(l)},
\end{equation}
where $p\wedge q=\min(p,q)$ and $\y^{(0)}=\y_i$. We also add axial points so that we can go outside of the convex hull of $\mathcal{M}_m$:
\begin{equation}\label{eq:axial}
\bm c_j = \left(\frac{1.5}{p\wedge q}\sum_{\substack{{l\neq j;}\\{0\leq l \leq (p\wedge q)}}} \y_i^{(l)}\right)-0.5 \y_i^{(j)},
\end{equation}
where $j=0,1,\dots,p\wedge q$. Note that if the output dimension $q$ is larger than the input dimension $p$, the output space would be a $p$ dimensional manifold in a $q$ dimensional  space. Therefore, it is more natural to consider a simplex of the lower dimension. Implicit in this argument is the assumption that the input variables are all active, otherwise, the manifold dimension can be even lower. The second part $\mathcal{A}_2$ is generated by finding the midpoints between each $\y_i \in \mathcal{M}_m$ and its $k_1$-nearest neighbors  $N^{k_1}(\y_i)$. These points reside on the (extended) boundaries of the Voronoi cells and have equal distances to the end points. We choose $k_1=2(p\wedge q)$ by default. The last part $\mathcal{A}_3$ consists of points in $p\wedge q$-dimensional balls around each design output. The rational to use $p\wedge q$-dimensional balls is similar: if $q>p$, we should not generate the approximating points by $q$-dimensional balls since most of the points would fall outside the manifold. Instead we extract the tangent spaces around each design outputs using the simplexes constructed in the first part and generate uniform points in $p$-dimensional balls on the tangent spaces. This procedure is presented as Algorithm \ref{alg:aux} in the supplementary material and illustrated by an example in Figure \ref{fig:filldist}. We can see that for each output point $\y_i$, $d_i = \max_{a\in V_i^{out}\cap \mathcal{A}}d_y(\bm{a},\y_i)$ is a reasonable approximation of the exact local fill distance $h_i=\max_{a\in V_i^{out}}d_y(\bm{a},\y_i)$.

\subsubsection{Perturbation}
\begin{figure}
    \centering
    \includegraphics[width=0.4\textwidth]{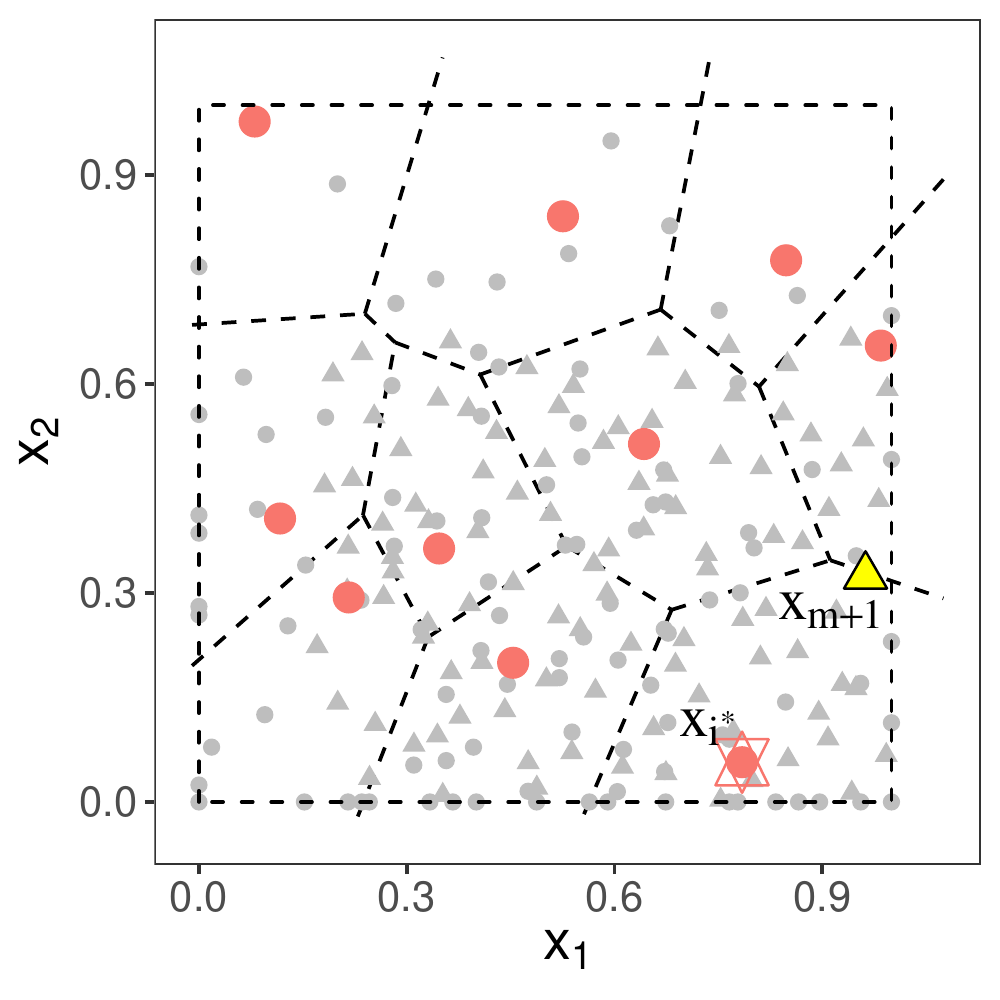}
    \caption{Illustration of the perturbation step. Red points are design points in the input space. The red star is the design points $\x_{i^*}$ corresponding to the output point $\y_{i^*}$ with the largest local fill distance. The yellow triangle is the next design point $\x_{m+1}$ chosen from the candidate set. Uniform candidates are denoted as the gray points and those candidates generated in the balls around $\x_{i^*}$ and its neighbor are denoted as gray triangles.}
    \label{fig:perturb}
\end{figure}

As mentioned earlier, we will perturb the input design point $\bm x_{i^*}$, corresponding to $\bm y_{i^*}$, the point in the output space with the largest local fill distance. The only thing we need to decide is how to perturb the $\bm x_{i^*}$. Ideally, we would like to find a new input point in such a way that the output will minimize the largest gap. However, this cannot be done optimally because $\f(\cdot)$ is expensive to evaluate. Therefore, we propose to perturb $\bm x_{i^*}$ to its maximum permissible level, which will also promote the space-fillingness in the input space. Thus, we choose the next point as the furthest point in the Voronoi cell $V^{in}_{i^*}$ of $\x_{i^*}$ as shown in Figure \ref{fig:perturb}, where
\[V^{in}_i=\{\x\in \mathcal{X}, d_x(\x,\x_i)\leq d_x(\x,\x_j),\;  j\neq i\}.\]

The foregoing computation is done as follows. A candidate set $\mathcal{C}$ is first built by uniform samples in a hypercube around $\x_{i^*}$. We then find the $k_2-$nearest neighbors $N^{k_2}(\x_{i^*})$ of $\x_{i^*}$, where we choose $k_2=2p$ by default so that there would be two neighbors on each dimension on average if the $\{\x_i\}_{i=1}^m$ were uniformly distributed. The set $\mathcal{C}$ is then augmented with uniform points in balls centered at $\x_{i^*}$ and its neighbors $N^{k_2}(\x_{i^*})$. The next design point $\x_{m+1}$ is chosen as
\begin{equation}
    \x_{m+1} \in \argmax_{ \x\in V_{i^*}^{in}\cap\mathcal{C}} d_x(\x,\x_{i^*}).
\end{equation}
This step and its parameter specification are detailed in Algorithm \ref{alg:greedy-perturb} in the supplementary material. The whole output space-filling design algorithm is summarized in Algorithm \ref{alg:OSFD} in the supplementary material. An illustration of the OSFD algorithm on the inverse radius function is shown in the middle panels of Figure \ref{fig:intro}. We can see that it does a good job filling the output space.

Now consider the following challenging exponential function $\f^{\alpha}_{\text{exp}}:[0,1]^2\to\Real^3$ taken from \cite{rhee2017space}:
\begin{equation}\label{eq:exp}
    \f^{\alpha}_{\text{exp}}(x_1,x_2)=\left(e^{-\alpha x_1}+e^{-\alpha x_2},e^{-2\alpha x_1}+e^{-2\alpha x_2},e^{-4\alpha x_1}+e^{-4\alpha x_2}\right).
\end{equation}

\begin{figure}
    \centering
    \includegraphics[width=0.7\textwidth]{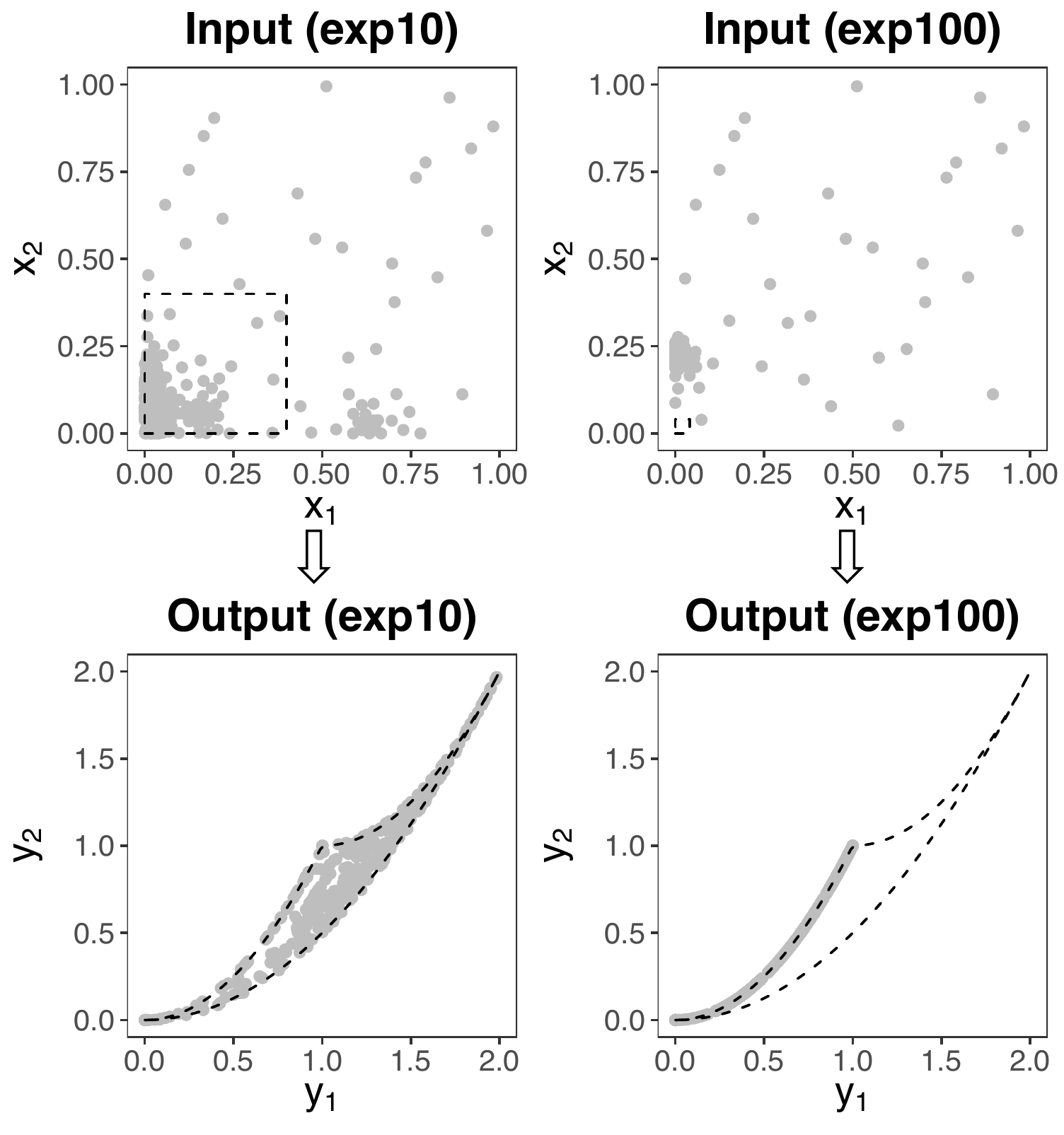}
    \caption{ Top row: 300 design points generated by the OSFD-greedy algorithm for the exponential function in \eqref{eq:exp} with $\alpha=10$ (left) and $\alpha=100$ (right). Bottom row: the corresponding output points projected to the first two coordinates. The intial design is generated by random LHD of size 30. The active region in the input space is shown as a box using dashed lines. The true output space is also shown using dashed lines.}
    \label{fig:greedy_exp}
\end{figure}

The parameter $\alpha$ controls the gradient of $\f^{\alpha}_{\text{exp}}$ in the input space $[0,1]^2$. When $\alpha$ is large, the gradient is large only in a very small area near the origin, making the active region difficult to locate. Consider $\alpha=10$ and $\alpha=100$. We can see from Figure \ref{fig:greedy_exp} that the proposed algorithm performs well for $\alpha=10$ but fails to cover the output space for $\alpha=100$. When $\alpha=100$, this function has large variation only in the region around $[0,0.04]^2$ (Figure \ref{fig:greedy_exp}: right panel), which means $99.84\%$ of the input design space will give almost identical responses. It is therefore almost impossible to find such a small area in the initial design. Moreover, at the initial stage, the approximation of the output space is inaccurate because the existing design points are far from the region $[0,0.04]^2$. Thus, the algorithm can get stuck in a local region. In the next subsection, we propose an improved algorithm that helps to jump out of the local regions.



\subsection{An Improved Algorithm}

Bayesian optimization \citep{garnett_bayesoptbook_2023} is a popular technique for the global optimization of expensive black-box functions. The key idea in Bayesian optimization is to introduce an acquisition function that includes not only the function value but also its uncertainty estimate. Expected improvement (EI) criterion \citep{jones1998efficient} is one such acquisition function. The EI criterion encourages the design points to explore the experimental region while exploiting the function, which aids in jumping out of local regions and enable the design points to move towards the global optimum. The EI algorithm uses Gaussian process (GP) modeling, which automatically gives the uncertainty estimates alongside predictions. However, as mentioned in the introduction, the high training cost of GP models can become a computational bottleneck.


It is well-known that a minimax design is based on a nearest neighbor predictor \citep{joseph2006limit}.  A nearest neighbor predictor is extremely fast. Its estimation can be done in $O(pm\log m)$ operations and prediction on $N$ points in $O(N\log m)$ operations, which are much smaller than the $O(m^3p)$ and $O(N^2)$ operations needed for a GP model. Unfortunately, nearest neighbor predictor is not based on a stochastic model and therefore, it does not come with uncertainty estimates as in GP modeling. Thus, the nearest neighbor approach is possible only if we can develop an uncertainty estimate. 


Let $h(\x)$ denote the local fill distance in the output space at an input value $\x$. Thus, given the data $\bm h=(h_1,\ldots,h_m)'$,  for a nearest neighbor predictor
\[\Expect\{h(\x)|\bm h\}=h_i\;\;\textrm{for}\;\; \x\in V_i^{in}(\x).\]
Motivated by the Brownian random fields \citep{zhang2014fractional}, we postulate a variance for the nearest neighbor predictor to be
\[\Var\{h(\x)|\bm h\}=\sigma^2||\x-\x_i||\;\;\textrm{for}\;\; \x\in V_i^{in}(\x).\]
This has the desirable property that $\Var\{h(\x_i)\}=0$ for $i=1,\ldots,m$ and that the variance increases as the prediction point moves away from the design points, just like the posterior variance in a Gaussian process model. Assuming normality, we have
\begin{equation}
    h(\x)|\bm h\sim \mathcal{N}\left(h_i,\sigma^2||\x-\x_i||\right) \;\;\textrm{for}\;\; \x\in V_i^{in}(\x).
\end{equation}
%
Thus, the expected improvement acquisition function can be obtained as
\begin{align}
    \text{EI}(\x;\mathcal{D}_m) & = \Expect \left[\max{(0,h(\x)-h_{max})}|\bm h\right]\nonumber\\
    & = s(\x) {\{u(\x)\Phi((u(\x))+\phi(u(\x))\}}\;\;\textrm{for}\;\; \x\in V_i^{in}(\x), 
\end{align}
where $h_{max} = \max_i {h_i}$, $s(\x)=\sqrt{\sigma^2\|\x-\x_i\|}$, $u(\x)=(h(\x)-h_{max})/s(\x)$, $\phi(\cdot)$ is the density function of the standard normal random variable and $\Phi(\cdot)$ is its cumulative distribution function. An estimate of $\sigma^2$ can be obtained by maximizing the leave-one-out cross-validation likelihood as \citep{geisser1979predictive}: 
\begin{equation}\label{eq:sigma}
    \widehat{\sigma}^2 = \frac{1}{m} \sum_{i=1}^m\frac{\left(h_i-h(\x_i^{(1)})\right)^2}{\|\x_i-\x_i^{(1)}\|},
\end{equation}
where $\x_i^{(1)}=\argmin_{\x\in\mathcal{D}_m\setminus\{\x_i\}} ||\x_i-\x||$. 

As before, we generate a candidate set of points $\mathcal{C}$ and obtain the next design point as:
\begin{equation}\label{eq:OSFD-EI}
    \x_{m+1} = \argmax_{\x \in\mathcal{C}} \text{EI}(\x;\mathcal{D}_m).
\end{equation}
To promote exploring across the whole input space, the candidate set $\mathcal{C}$ includes uniform random points in the hypercube $[0,1]^p$ instead of a local region around $
\x_{i^*}$. The new algorithm is shown in Algorithm \ref{alg:ei-perturb} in the supplementary material. To distinguish this design from the previous greedy strategy, we will refer to the new design as OSFD-EI and the previous design as OSFD-greedy.

\begin{figure}[h]
    \centering
    \includegraphics[width=0.7\textwidth]{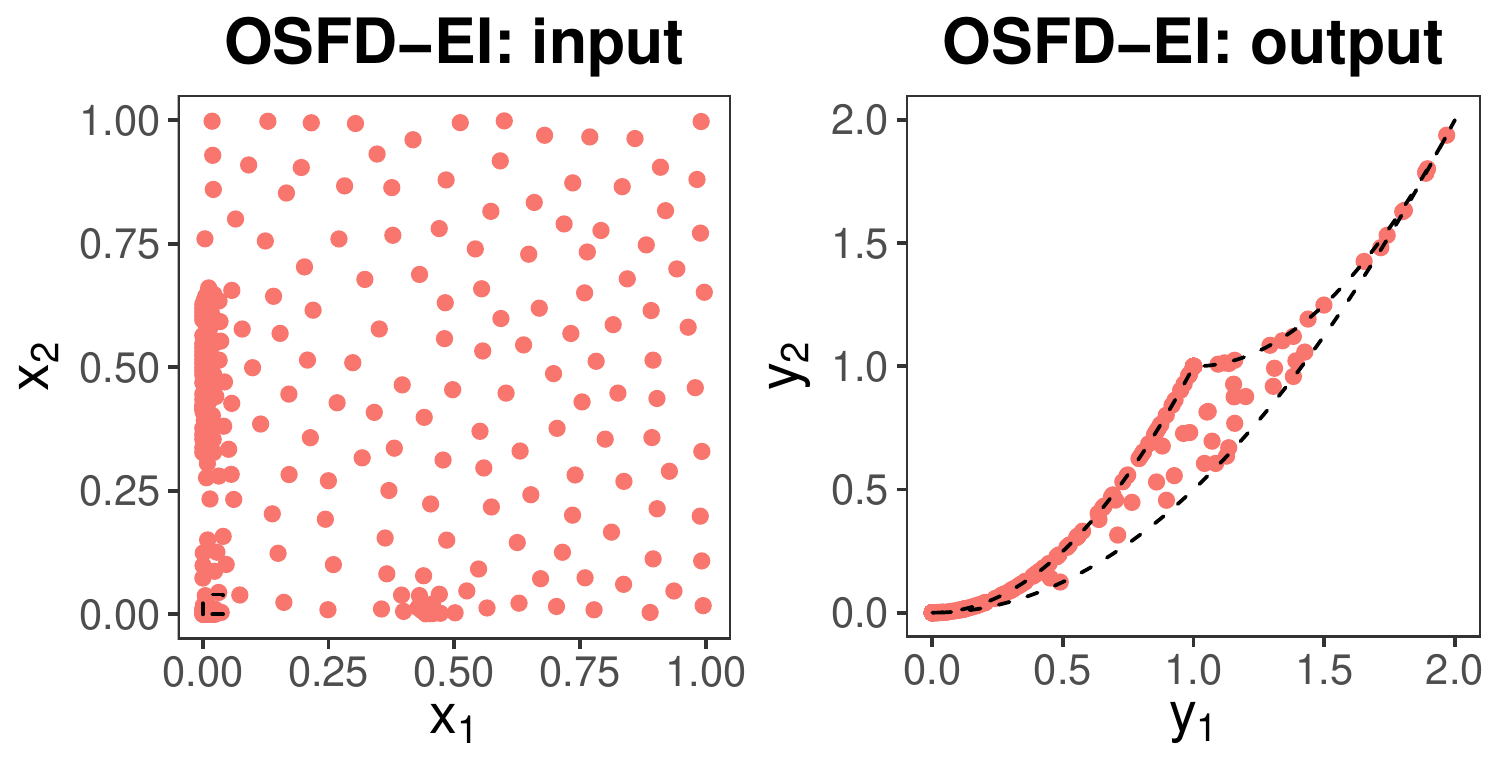}
    \caption{Design points and outputs for $\f^{100}_{exp}$ generated by OSFD-EI.}
    \label{fig:exp}
\end{figure}

The 300 points generated using the OSFD-EI algorithm for the exponential function with $\alpha=100$ is shown in Figure \ref{fig:exp}. The initial design is random LHD of size 30. We can see that the new design tends to explore the input space like an ISFD and is able to jump out of the local regions and fill the output space reasonably well. On the other hand, the approach in \cite{rhee2017space} required thousands of function evaluations to get a similar result.

\begin{figure}[h]
    \centering
    \includegraphics[width=0.7\textwidth]{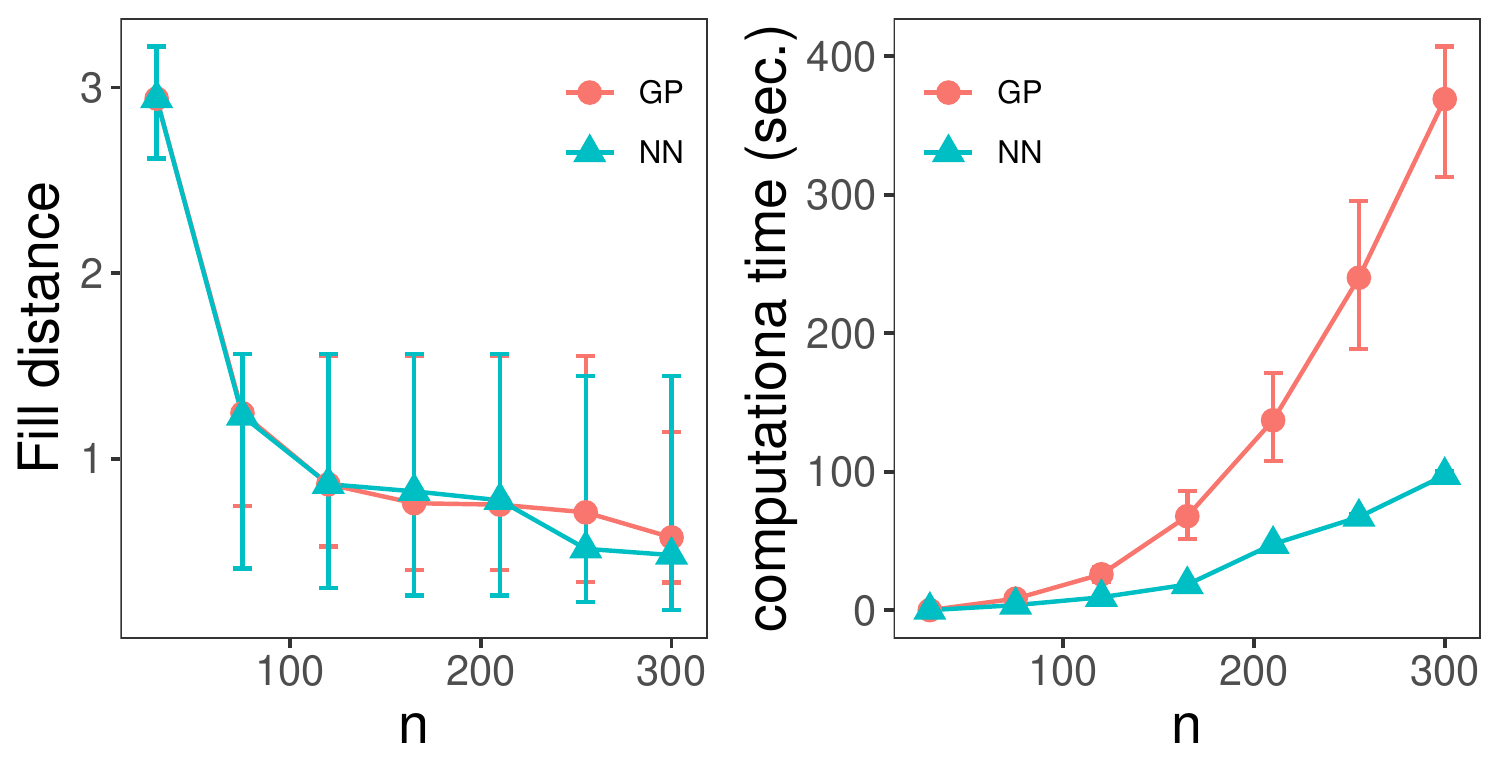}
    \caption{The fill distance and computational time for an OSFD of size $n$ generated by nearest neighbor-based OSFD-EI and GP-based OSFD-EI. The underlying function is $\f^{100}_{exp}$. The lines denote the average over 10 replications and the error bars represent the 5th and 95th quantiles.}
    \label{fig:compare_nn_gp}
\end{figure}

We also tried a GP model instead of the nearest neighbor (NN) predictor on the exponential function with $\alpha=100$. The GP model was fitted using the R package \texttt{DiceKriging} \citep{roustant2012dicekriging} at the default settings.  The fill distance (left) and computational time (right) averaged over 10 replications are shown in Figure \ref{fig:compare_nn_gp} for various values of $n$. As expected, the NN predictor makes the algorithm run much faster than with GP. The fill distance is  comparable  to that of the GP model.


\section{Simulations}\label{results}

\begin{figure}
    \centering
    \includegraphics[width=0.4\textwidth]{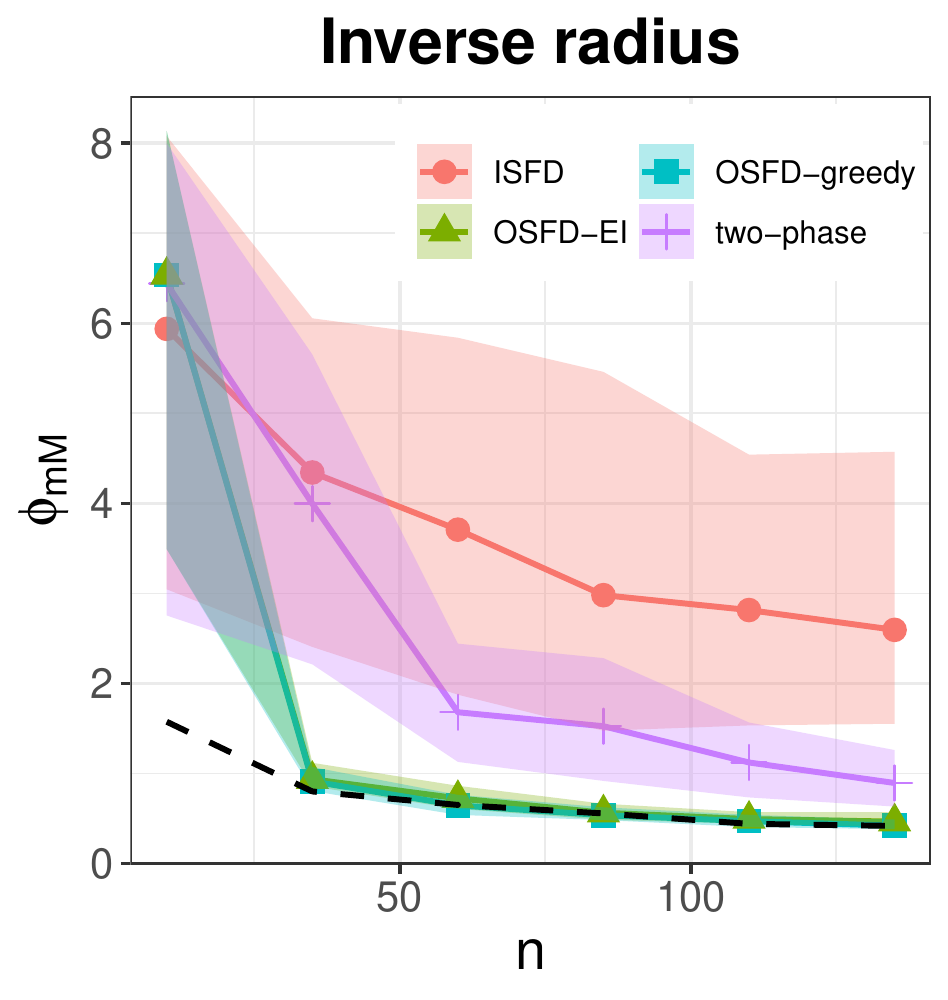}
    \caption{Fill distance for OSFD-EI, OSFD-greedy, ISFD, two-phase, and optimal minimax OSFD (black dashed line) against run size for the inverse radius function in  \eqref{eq:ir}. Lines denote the average values and the shaded bands mark the 5th and 95th quantiles.}
    \label{fig:compare_ir}
\end{figure}

In this section, we investigate the performance of the two proposed OSFD algorithms. We consider three test functions: the inverse-radius function $\f_{\text{ir}}$ \eqref{eq:ir}, the exponential function $\f^\alpha_{\text{exp}}$ \eqref{eq:exp} with different $\alpha$, and a modified Easom function \citep{solteiro2010particle}: 
\begin{equation} \label{eq:linear}
    f^p_{\text{esm}}(\x)=\prod_{i=1}^p \cos(2\pi x_i)\exp\left(-\frac{\pi^2(2x_i-1)^2}{p}\right)
\end{equation}
with different input dimension $p$. We will use the fill distance in the output space to quantify the performance. A two-phase algorithm adapted from \cite{lu2021input} is also included for comparison. In the first phase of this algorithm, we use a random LHD of size $n/4$ (rounded to the nearest integer) to build a multivariate GP model. In the second phase, we predict the responses of $100n$ uniform random inputs using this GP model and choose $3n/4$ predicted responses based on maximin criterion. The corresponding input points are chosen as the remaining $3n/4$  points of the $n$-point design. Clearly, the performance of this algorithm depends largely on how well the GP model fits the underlying mapping $\f$ in the first phase. We used \texttt{DiceKriging} \citep{roustant2012dicekriging} for fitting the GP model and the R package \texttt{maximin} \citep{rpackage_maximin} for finding the maximin points.

Figure \ref{fig:compare_ir} shows the simulation results for the inverse-radius function. We initialize the OSFD by a random LHD with sample size $n_0=10$ and replicate the simulation 20 times. The solid line indicates the mean value of the fill distance and the shaded band represent the 90\% confidence intervals. For each $n$, we also generate an ISFD using random LHD and compute its fill distance in the output space. We can see that both OSFD-greedy and OSFD-EI algorithms outperform the ISFD and the two-phase algorithm. In fact, both the proposed algorithms quickly attain the optimal fill distance after a few steps. The optimal value is obtained by directly running mMc-PSO algorithm using the R package \texttt{minimaxdesign} \citep{mak2016minimaxdesign} on the true output space. Another advantage of our algorithms compared to the two-phase algorithm is its speed. To generate a design of size 150, the computational time for OSFD-EI and OSFD-greedy are $1.1$s and $0.7$s respectively while it takes $6.8$s for the two-phase algorithm in a 2.6 GHz 6-Core Intel Core i7 processor. This computational saving becomes even more substantial as  the design size increases.

\begin{figure}
    \centering
    \includegraphics[width=1\textwidth]{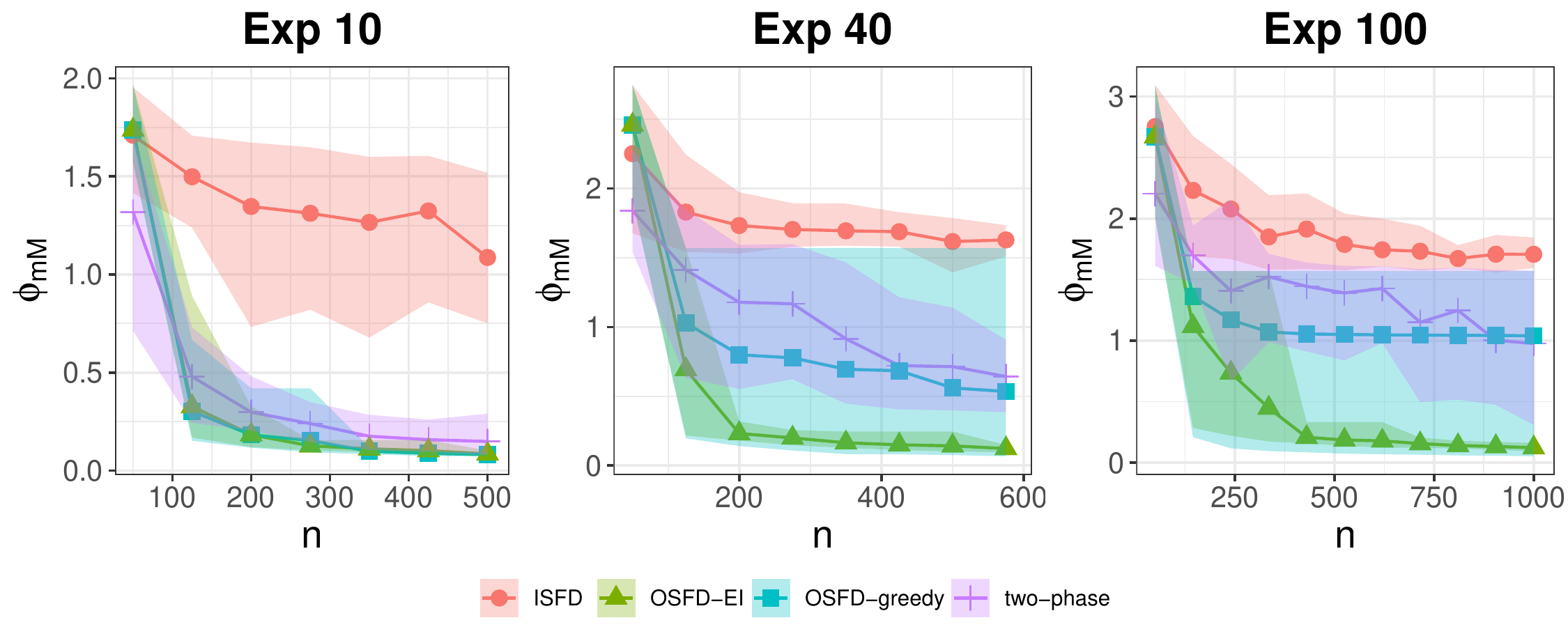}
    \caption{Fill distance for OSFD-EI, OSFD-greedy, two-phase, and random LHD against run size for the exponential function in \eqref{eq:exp} with $\alpha=10, 40, 100$. Lines denote the average values and the shaded bands mark the 5th and 95th quantiles.}
    \label{fig:compare_exp}
\end{figure}

The foregoing simulation is repeated on the exponential function in (\ref{eq:exp}) with $\alpha=10,40,100$. We initialize the OSFD algorithms using $n_0=50$ random LHD points in the input space. We can see from Figure \ref{fig:compare_exp} that both the OSFD algorithms are superior to ISFD (using random LHD) when $\alpha=10$. As the $\alpha$ increases, the problem becomes more challenging. We can see that the OSFD-EI performs much better than the OSFD-greedy and the two-phase algorithm for large $\alpha$.


\begin{figure}
    \centering
    \includegraphics[width=1\textwidth]{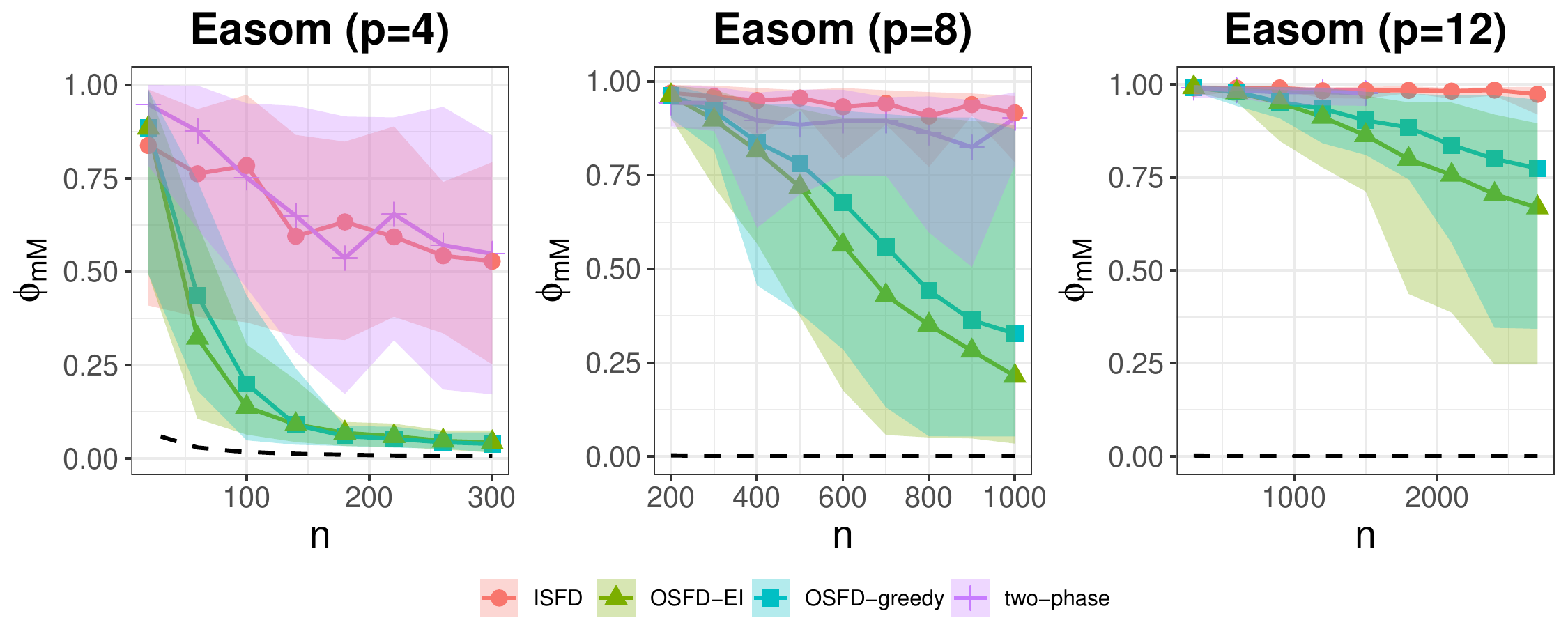}
    \caption{Fill distance for OSFD-EI, OSFD-greedy, random LHD, two-phase, and optimal minimax OSFD (black dashed line) against run size for the modified Easom function  with $p = 4,8,12$ with initial design size of 20, 200, 300. Lines denote the average values and the shaded bands mark the 5th and 95th quantiles. The two-phase algorithm for $p=12$ is terminated at  $n=1500$ due to high computational time and memory requirement.}
    \label{fig:compare_linear}
\end{figure}

Finally, we investigate the impact of input dimension on the performance by considering the modified Easom function in (\ref{eq:linear}). Figure \ref{fig:compare_linear} shows that OSFD consistently outperforms ISFD (using random LHD) and the two-phase algorithm, however, the advantage of OSFD diminishes as the dimension increases. This is not unexpected as high dimension makes any feasible set of design points sparse in the input space.  Improving the performance of OSFD for high dimensional problems could be an important topic for future research.



\section{Applications}\label{application}
In this section, we will present two applications of output space-filling design.
\subsection{Inverse design}
In the inverse design problem, the goal is to provide a suitable input that can produce a desired output within a reasonable degree of accuracy. This output can be crystal properties \citep{ren2022invertible}, modulating properties of optical devices \citep{molesky2018inverse}, or the acoustic properties of material structures \citep{krishna2022inverse}, etc. If  a large number of targets are of interest, OSFD design can quickly give the set of inputs that can approximately achieve the targets. 

Here we consider a simple example in which we hope to control a robot arm  in a two-dimensional plane \citep{an2001quasi}. The robot arm has four extendable segments of lengths $L_1, L_2, L_3, L_4$ and are at angle $\theta_1,\theta_2, \theta_3,\theta_4$ to the horizontal coordinate axis of the plane. The location of the end of the robot arm $(u,v)$ is:
\begin{align*}
    u = \sum_{i=1}^4 L_i \cos{\left(\sum_{j=1}^i\theta_j\right)},\quad
    v = \sum_{i=1}^4 L_i \sin{\left(\sum_{j=1}^i\theta_j\right)},
\end{align*}
where $L_i\in[0,1]$ and $\theta_i\in[0,2\pi]$ for $i=1,2,3,4$. Clearly, the arm's range of motion is within a circle of radius $4$. 

\begin{figure}[h]
\begin{subfigure}[c]{.5\textwidth}
    \centering
    \includegraphics[width=0.9\textwidth]{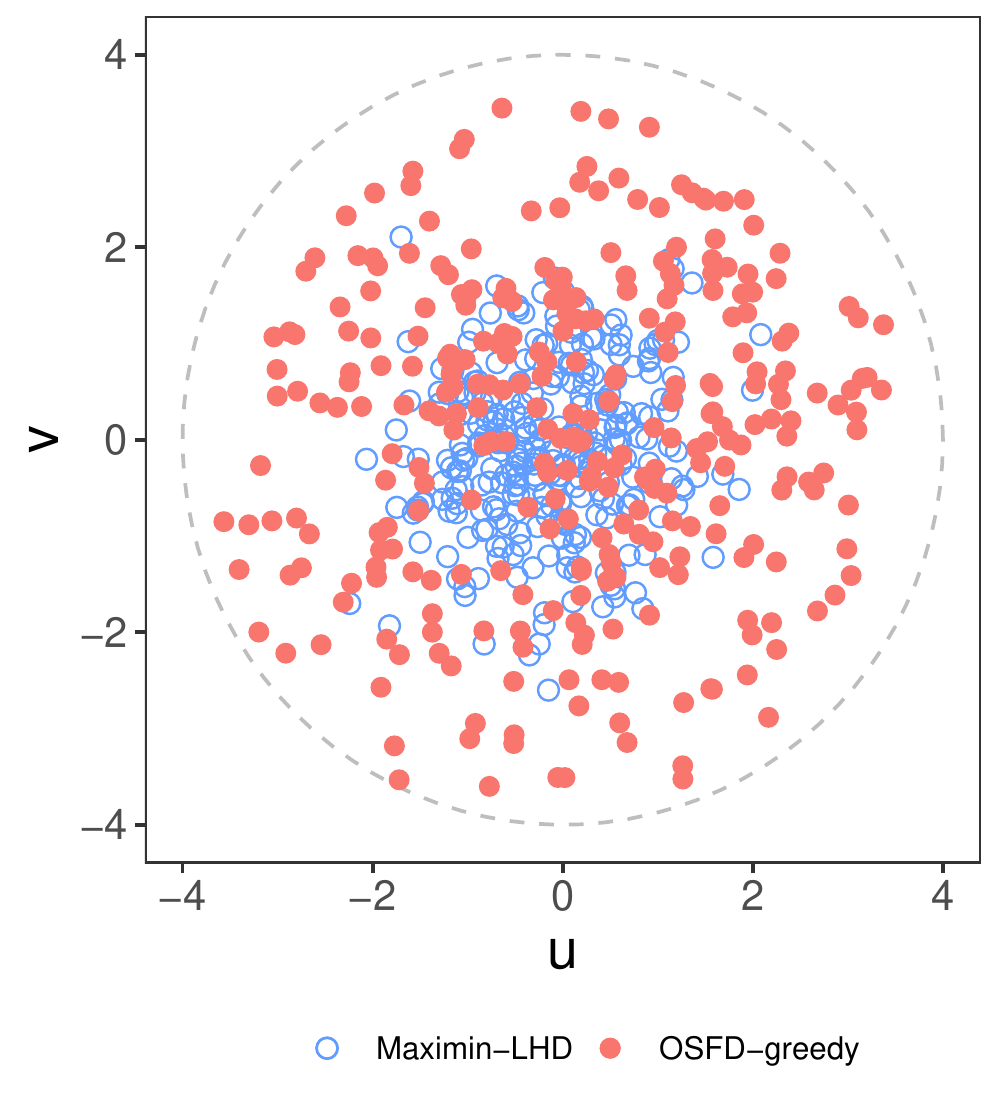}
    \caption{}
    \label{fig:robot_scatter}
\end{subfigure}
\begin{subfigure}[c]{.5\textwidth}
    \centering
    \includegraphics[width=0.9\textwidth]{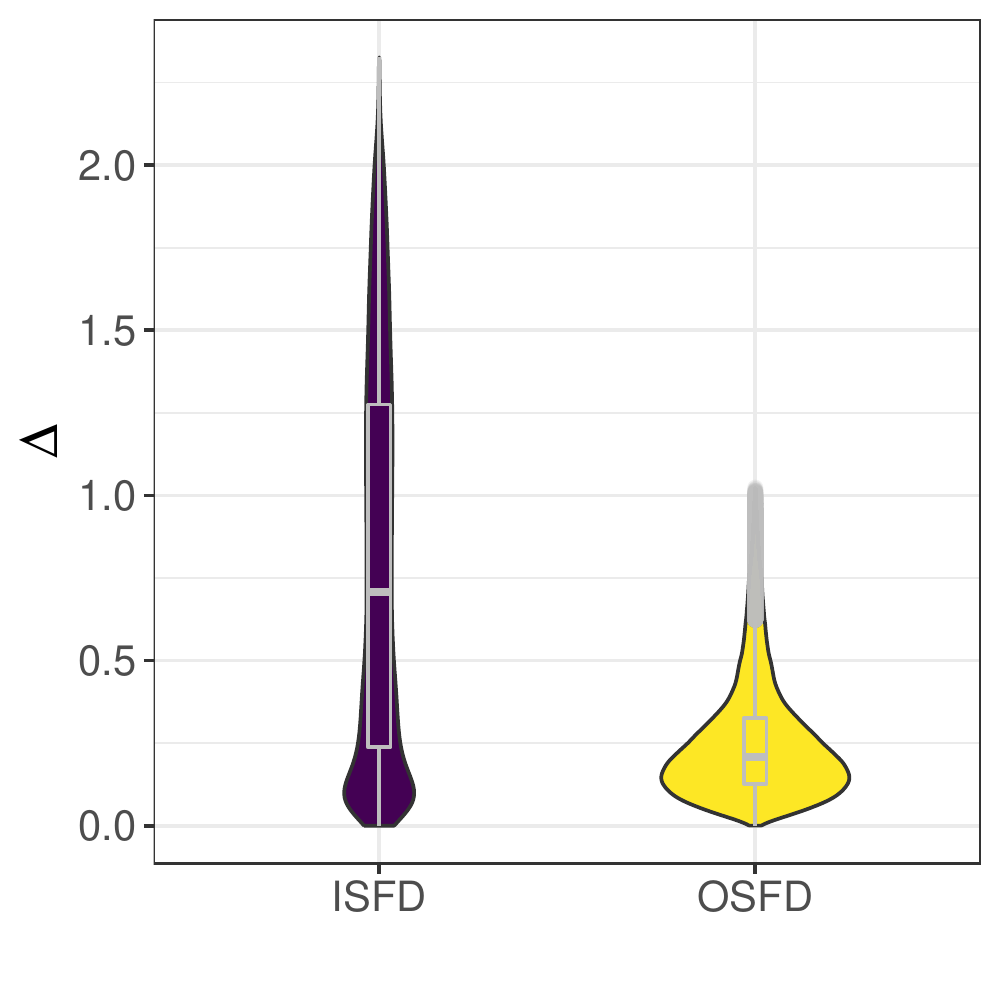}
    \caption{}
    \label{fig:robot_error}
\end{subfigure}
    \caption{(a) Scatter plot of the outputs from OSFD and Maximin-LHD; (b) The distances from the targets to the closest  output point ($\Delta$). Total number of targets is 100,030. Design size is 300. Initial design is generated by maximin LHD of size 30.}
\end{figure}

Figure \ref{fig:robot_scatter} shows that the  outputs from the OSFD-greedy has a considerably better coverage of the full output space than the outputs generated by an ISFD (generated using the maximin LHD). To quantify the approximation error in an inverse design application, we generate 100,030 uniform points inside the circle shown in  Figure \ref{fig:robot_scatter} (left). The distances from each target $y^*$ to the closest output point  $\Delta = d_y(y^*,N^1(y^*))$ should be as small as possible. The nearest output point $N^1(y^*)$ is found from the outputs of OSFD and ISFD, respectively. From Figure \ref{fig:robot_error}, we can see that the distances from the target outputs produced from OSFD are in general significantly lower than those from the ISFD.

\subsection{Feature-based modeling}
In this section, we present an exemplar case study of feature-based modeling within the field of material informatics based on \cite{generale2021reduced} in which a model-based linkage between virtually generated 5-harness satin (5HS) ceramic matrix composite (CMC) microstructures and their effective orthotropic thermal conductivity was developed. A GP-based model was trained through an active learning framework to reduce the computational burden inherent in performing finite element (FE) based thermal analyses. Subsequent microstructures selected for evaluation were identified through maximum posterior uncertainty and constrained to a preexisting microstructure ensemble. The input features for this model were extracted from n-point spatial correlations of the microstructure (also referred to as n-point statistics) \citep{torquato_random_2002}. As a collective, n-point spatial correlations provide a hierarchy of increasingly complex descriptions of the material microstructure. 1-point spatial correlations capture the probabilities of finding a specific local material state (i.e., microscale constituent) at any randomly selected voxel in a discretized representative volume element (RVE), more commonly referred to as the local material state's volume fraction. With increasing complexity, 2-point spatial correlations define the probability of finding two specified local states at the head and tail of a randomly placed vector in the RVE. From solely the description of the first two n-point spatial correlations, it quickly becomes apparent that this representation of a materials microstructure is extremely high-dimensional in nature, with each dimension capturing a singular statistic related to the spatial arrangement of local states \citep{kalidindi2015hierarchical}. In this case study, 2-point spatial correlations were computed for an ensemble of virtually generated 5HS CMC RVEs, with principal component analysis (PCA) performed to extract a low-dimensional representation of microstructure. Each microstructure in this ensemble was generated through the open-source software package TexGen \citep{lin_modelling_2011}, with five generating geometric parameters considered, as listed in Table \ref{table:CMC}.

\begin{table}
\centering
\caption{Geometric parameters for CMC}
\label{table:CMC}
\begin{tabular}{lccc}
\hline
Microstructural  Dimension & Minimum & Nominal & Maximum \\
\hline
Tow Major Axis - $t_w$         & 788     & 985     & 1182    \\
Tow Minor Axis - $t_h$           & 92      & 115     & 138     \\
Tow Spacing - $t_s$             & 880     & 1100    & 1320    \\
Ply Spacing - $p_c$            & 257     & 275     & 292     \\
Matrix Thickness - $m_t$        & 47      & 78      & 109    \\
\hline
\end{tabular}
\end{table}

\begin{figure}
    \centering
    \includegraphics[width=1\textwidth]{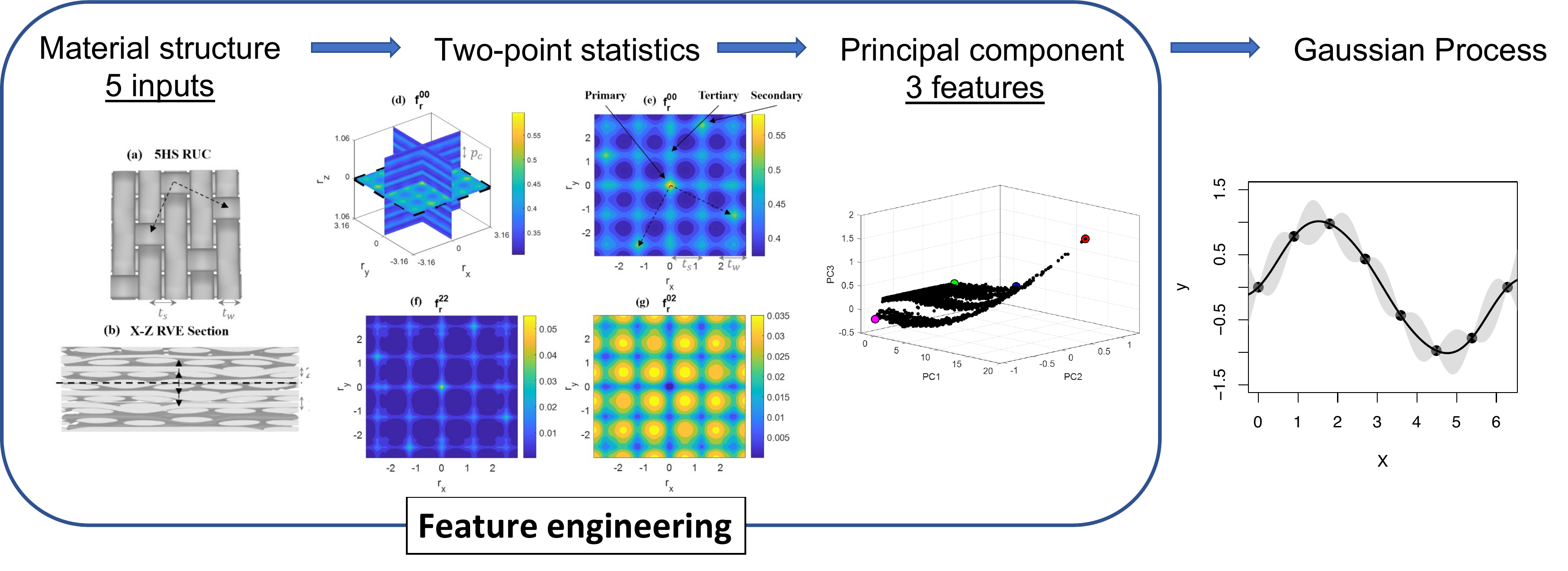}
    \caption{The process to build a reduced-order model for CMC thermal conductivity. Adapted from \cite{generale2021reduced}.}
    \label{fig:CMC}
\end{figure}

The first three geometric parameters listed controlled the generation of each ply within the RVE, defining the cross-sectional shape of the reinforcing tows through the tow major and minor axes, and the tow spacing within the woven architecture. The RVE was then assembled through the stacking of eight plies of the 5HS repeating unit cell (RUC) \citep{deo_analysis_1996}. The ply spacing then defined the distance from ply to ply within this stack. With these four geometric parameters, completely dense voxelated microstructures, consisting of tow or matrix voxels, were output by TexGen \citep{lin_modelling_2011} with $100^3$ total voxel count. Subsequently, matrix voxels were reassigned to pores by a threshold defined by the matrix thickness parameter, resulting in RVEs with three constituents. A Maximin LHD was then used to generate 3,125 unique microstructures. It should be highlighted that for a microstructure of size $100^3$ with three constituents, the collection of 2-point spatial correlations results in $3\times100^3$ dimensions, including two sets of auto-correlations, and one set of cross-correlations. A schematic demonstrating the overall process employed in their work can be seen in Figure \ref{fig:CMC}, which outlines the feature engineering protocol from microstructure ensemble generation, to computation of 2-point spatial correlations and through performing PCA to establish three-dimensional inputs into the predictive GP model. 
While it was demonstrated that the active learning framework significantly reduced the computational demand, a core limitation consisted of defining the geometric parameterized microstructure input space to be space-filling rather than the low-dimensional representation (i.e. the final input to the GP model), leading to suboptimal model building as coverage of the output space, in this case the principal components of the set of 2-point spatial correlations, were poorly clustered. Subsequently, we present the benefits of employing an OFSD on their three-dimensional microstructure dataset. 

\begin{figure}
    \centering
    \includegraphics[width=1\textwidth]{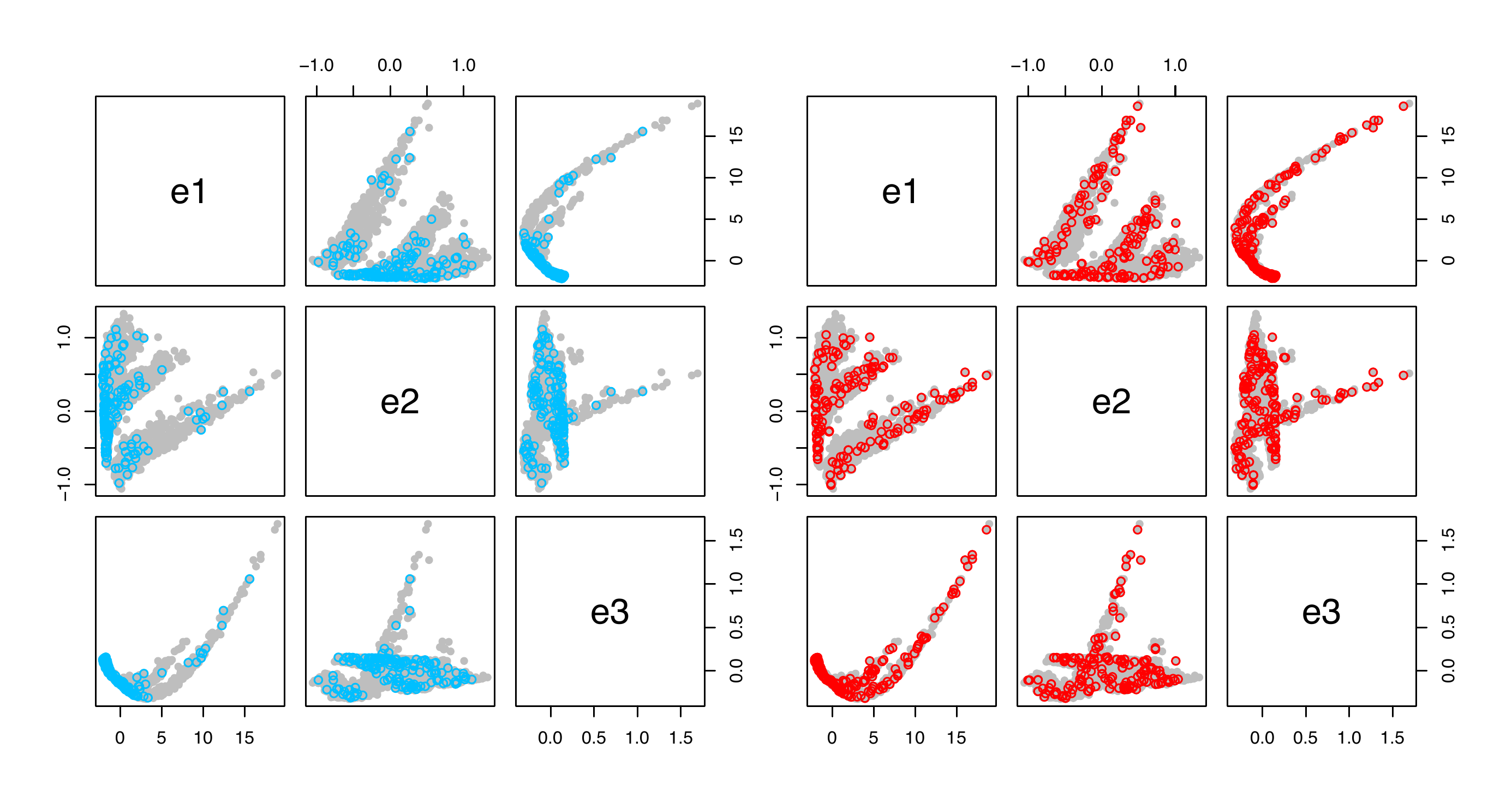}
    \caption{Gray, blue, red points are features from the whole dataset, maximin LHD and OSFD-EI, respectively.}
    \label{fig:cmc_scatter}
\end{figure}

The feature engineering protocol shown in Figure \ref{fig:CMC}, can be described as a mapping $\f(\x): \mathcal{X}\to\mathcal{E}$, where $\X$ is the input space of $(t_w,t_h,t_s,p_c,m_t)$ and $\mathcal{E}$ is the feature space of the three major features $(e_1,e_2,e_3)$. 
In order to demonstrate the proposed utility of employing the OSFD algorithm on this dataset, 50 microstructures from the complete ensemble were selected as an initialization subset to estimate the principal component (PC) basis and lock it in place with sequential selection of microstructures in the feature space. The OSFD-EI algorithm was then applied to sequentially select an additional 100 microstructures from the ensemble best filling this feature space, with the results displayed in Figure \ref{fig:cmc_scatter}. The identified OSFD is shown overlaid against the PCA representation of the complete ensemble in gray, demonstrating the impressive coverage of this input feature space through a minimal collection of all available microstructures. For comparison purposes, the left panel of Figure \ref{fig:cmc_scatter} displays the results from selecting 150 microstructures utilizing an ISFD, as generated through a 150-run Maximum LHD and selecting the nearest neighboring point. Quantitatively, the fill distance for the features generated by ISFD is $3.43$ while that produced by OSFD is 0.67. This direct comparison clearly displays the poor coverage offered through selecting model-building inputs with an ISFD in comparison to the proposed OSFD for this application.

There are several benefits of employing an OSFD instead of an ISFD in this application. Most importantly, the coverage provided through 150 points generated through the OSFD algorithm is nearly as good as the complete ensemble of 3,125 points, conventionally generated through an ISFD. This fact has important implications for the overall cost of training the final GP-based structure-property linkage, as microstructure generation and the application of PCA to the complete ensemble with size $3,125\times 3\times100^3$ is computationally expensive and represents overhead which can be substantially reduced through the use of an OSFD. The use of an OSFD could also be leveraged to generate additional structures in regions of the input space with insufficient coverage, clearly visualized in Figure \ref{fig:cmc_scatter}. The use of such additional points may lead to a more robust final model, as GP-based models are well known to extrapolate poorly.

\section{Conclusions}
\label{conclusion}


It is common to use a space-filling design in the input space to generate the computer model outputs and develop the input-output relationship. In this article we have demonstrated that for several applications, and contrary to conventional model building workflows, filling the output space is more desirable than filling the input space. We have proposed a sequential design that identifies the largest gap in the output space and generates an input point to fill-in that gap. Two versions of the sequential design are proposed: a greedy algorithm and an expected improvement-based algorithm. They are fast and model-independent, and therefore, we hope that they will have broad applications. We have demonstrated the usefulness of the proposed method on two applications involving inverse design and feature-based modeling.

Although we have used nearest neighbor method in our sequential design algorithms because of its computational speed, other surrogate modeling techniques can be used as long as they can be trained quickly with large amounts of data. The traditional GP model did not perform well in our applications, but we believe nonstationary GP models that scale well with the size of data can further improve the OSFD. We leave this as a topic for future research.





\bibliographystyle{asa}
\bibliography{Bibliography.bib}
\clearpage
\pagebreak
\begin{center}
\textbf{\large Supplemental Materials: Sequential Designs for Filling Output Spaces}
\end{center}

\setcounter{section}{0}
\renewcommand{\thesection}{S-\Roman{section}}

\section{Algorithms}

\begin{algorithm}[ht]
   \caption{\texttt{ApproxGen}: approximating points generation}
   \label{alg:aux}
   \begin{algorithmic}[1]
   \Require Output points $\mathcal{M}_m$, input dimension $p$, output dimension $q$.
   \State \textbf{Initialization:} approximating point set $\mathcal{A}\leftarrow\mathcal{A}_1\leftarrow\mathcal{A}_2\leftarrow\mathcal{A}_3\leftarrow\varnothing$, number of neighbors $k_1=2 (p\wedge q)$ considered in part $\mathcal{A}_2$.
        \For{$i=1,\dots,m$} 
            \State  Find the $(p\wedge q)$-nearest neighbors $N^{p\wedge q}(\y_i)$ of $\y_i$.\Comment{$\mathcal{A}_1$ construction}
            \State  Compute the centroid $\bm c$ of the simplex constructed by $\y_i$ and $N^{p\wedge q}(\y_i)$ (Eq. \ref{eq:centroid})
              and axial points $\bm c_0,\dots,\bm c_{p\wedge q}$ on the extended medians (Eq. \ref{eq:axial}):
                
            \State   Augment $\mathcal{A}_1=\mathcal{A}_1\cup \{\bm{c},\bm{c}_0, \dots, \bm {c}_{p\wedge q}\}$ .
            
            \State  Find the $k_1$-nearest neighbors $N_i^{k_1}=\{\y_i^{(l)}: l=1,\dots,k_1\}$ of $\y_i$.\Comment{$\mathcal{A}_2$ construction}
            \State  For each neighbor $\y_i^{(l)}$, compute its midpoint with $\y_i$
                \begin{equation*}
                \bm{m}_l = \frac{\y_i+\y_i^{(l)}}{2}.
                \end{equation*}
             \State   Augment $\mathcal{A}_2=\mathcal{A}_2\cup \{\bm{m}_1,\dots, \bm{m}_{k_1}\}$.

            \State The radius of ball:   $r_i = d_y(\y_i^{(1)},\y_i)$; \Comment{$\mathcal{A}_3$ construction}
            
            \algstore{myalg}
    \end{algorithmic}
\end{algorithm}

\begin{algorithm}[ht]
    \begin{algorithmic}
    \algrestore{myalg}
            \If{$p\geq q$} 
                \State Generate uniform points $\mathcal{B}_i$ of size $k_1+2(q+1)+1$ in a $q$ dimensional ball centered at $\y_i$ of radius $r_i$;
            \Else
                \State Perform principal component analysis (PCA) on the the set of points $\y_{i} \cup N^{p\wedge q}(\y_i)$ to obtain the tangent space at point $\y_{i}$ and generate uniform points $\mathcal{B}_i$ of size $k_1+2(p+1)+1$ in a $p$ dimensional ball centered at $\y_i$ of radius $r_i$ on the tangent space.
            \EndIf
            \State Augment $\mathcal{A}_3=\mathcal{A}_3\cup \mathcal{B}_i$. 
        \EndFor
        \State $\mathcal{A}=\mathcal{A}_1\cup\mathcal{A}_2\cup\mathcal{A}_3$.
        \State Remove repeated points in $\mathcal{A}$.
   \State \textbf{Return:} $\mathcal{A}$.
\end{algorithmic}
\end{algorithm}

\begin{algorithm}[ht]
   \caption{\texttt{filldistance}: Local fill distance for each design output}
   \label{alg:filldistance}
   \begin{algorithmic}[1]
   \Require Output points $\mathcal{M}_m$, input dimension $p$, output dimension $q$.
        \State Generate approximating points by $\mathcal{A}=$ \texttt{ApproxGen} ($\mathcal{M}_m,p,q$).
        \State Assign each point in $\mathcal{A}$ to its closest output point. 
        \For {$i=1,\dots,m$}
        \State Compute the local fill distance $d_i= \max_{a\in \mathcal{Z}_i}d_y(a,\y_i)$, where $\mathcal{Z}_i$ is the set of points in $\mathcal{A}$ assigned to $\y_i$.
        \EndFor
   \State \textbf{Return:} $\{d_i\}_{i=1}^m.$
\end{algorithmic}
\end{algorithm} 

\begin{algorithm}[ht]
   \caption{\texttt{greedy-perturbation}: greedy perturbation}
   \label{alg:greedy-perturb}
   \begin{algorithmic}[1]
   \Require Design $\mathcal{D}_m$, input dimension $p$, local fill distance $\{d_i\}_{i=1}^m$.
    \State Choose the perturbed point $\x_{i^*} $ (Eq. \ref{eq:istar} in the paper) and find its $k_2=2p$ nearest neighbors $N^{k_2}(\x_{i^*})$. 
    \State Generate $10p(k_2+1)$ scrambled Sobol sequence as the candidate set $\mathcal{C}$ within the hypercube $\bigotimes_{j=1}^{p}\left[0\vee \left(\x_{i^*}-d(\x_{i^*},\x_{i^*}^{(k_2)})\right)_j,1\wedge \left(\x_{i^*}+d(\x_{i^*},\x_{i^*}^{(k_2)})\right)_j\right]$.

    \For{$j=1,\dots,k_2$}
    \begin{equation*}
        r_j = d_x(\x_{i^*},\x_{i^*}^{(j)})
    \end{equation*}
    \State Augment $\mathcal{C}$ with uniform points of size $10p$ in balls centered at $\x_{i^*}^{(j)}$ with radius $r_i$.
    \EndFor
    \State Augment $\mathcal{C}$ with uniform points of size $10p$ in the ball centered at $\x_{i^*}$ with radius $d_x(\x_{i^*},\x_{i^*}^{(1)})$.
    
    \State Determine $\x_{m+1} \leftarrow \argmax_{\x\in\mathcal{C}\cap V^{in}_{i^*}} d_x(\x,\x_{i^*})$.
   \State \textbf{Return:} $\x_{m+1}.$
\end{algorithmic}
\end{algorithm}

\begin{algorithm}[ht]
   \caption{\texttt{EI-perturbation}: Expected improvement perturbation}
   \label{alg:ei-perturb}
   \begin{algorithmic}[1]
   \Require Design $\mathcal{D}_m$, input dimension $p$, local maximum distance $\{h_i^{(m)}\}_{i=1}^m$.
    \State Generate $10 m$ uniform random points as the candidate set $\mathcal{C}$ within the unit hypercube.
    \State Find $\x_{i^*} $ (Eq. \ref{eq:istar} in the paper) and its $k_2=2p$ nearest neighbors $N^{k_2}(\x_{i^*})$. 
    \For{$j=1,\dots,k_2$}
    \begin{equation*}
        r_j = d_x(\x_{i^*},\x_{i^*}^{(j)})
    \end{equation*}
    \State Augment $\mathcal{C}$ with uniform points of size $10p$ in balls centered at $\x_{i^*}^{(j)}$ with radius $r_i$.
    \State Augment $\mathcal{C}$ with the mid points between each point in $\mathcal{D}_m$ with its $j$th nearest neighbor.
    \EndFor
    \State Augment $\mathcal{C}$ with uniform points of size $10p$ in the ball centered at $\x_{i^*}$ with radius $d_x(\x_{i^*},\x_{i^*}^{(1)})$.
    
    \State Estimate $\sigma^2$ by  (Eq. \ref{eq:sigma} in the paper).
    \State Determine $\x_{m+1}$ by  (Eq. \ref{eq:OSFD-EI} in the paper).
   \State \textbf{Return:} $\x_{m+1}.$
\end{algorithmic}
\end{algorithm} 

\begin{algorithm}[ht]
   \caption{\texttt{OSFD}: Output space-filling design}
   \label{alg:OSFD}
   \begin{algorithmic}[1]
   \Require Black-box function $\f$, size of the design $n$, size of initial design $n_0$.
    \State Generate initial design $\mathcal{D}_0$ of size by maximin LHD and run computer experiment to get output $\mathcal{M}_{n_0}$.
    \For{$k=0:(n-n_0-1)$}
        \State Scale $\mathcal{M}_{n_0+k}$ to $[0,1]^q$ by the maximum and minimum value in each dimension. \Comment{(optional but suggested)}
        \State Run \texttt{filldistance}$(\mathcal{M}_{n_0+k},p,q)$ to get $\{d_i\}_{i=1}^{n_0+k-1}$.
        \State Run \texttt{greedy/EI-perturbation}$(\mathcal{D}_{n_0+k},p,\{d_i\}_{i=1}^{n_0+k-1})$ to determine $\x_{n_0+k+1}.$
        \State $\mathcal{D}_{n_0+k+1}\leftarrow\mathcal{D}_{n_0+k}\cup \{\x_{n_0+k+1}\};\quad \mathcal{M}_{n_0+k+1}\leftarrow\mathcal{M}_{n_0+k}\cup \{\f(\x_{n_0+k+1})\}$.
    \EndFor
   \State \textbf{Return:} $\mathcal{D}_n.$
\end{algorithmic}
\end{algorithm} 
\end{document}